\documentclass[pr,twocolumn,showpacs]{revtex4}
\usepackage{bm}
\usepackage{graphicx}
\usepackage{amsmath}
\usepackage{eufrak}
\usepackage{color}
\usepackage{upgreek}
\usepackage{subfigure}
\usepackage[unicode=true,colorlinks=true,citecolor=blue]{hyperref}
\usepackage{placeins}
\usepackage{lipsum}

\begin{document}

\title{Spin-dependent photogalvanic effects (A review)}

\author{E.L. Ivchenko$^1$  and  S.D. Ganichev$^2$}

\affiliation{$^1$ A.F. Ioffe Physico-Technical Institute, Russian
Academy of Sciences 194021 St. Petersburg, Russia  -  ivchenko@coherent.ioffe.ru}

\affiliation{$^2$ University of Regensburg,
Regensburg, 93040 Germany - sergey.ganichev@physik.uni-regensburg.de}

\begin{abstract}
In this paper we review both theoretical and experimental studies on spin-related photogalvanic effects. A short phenomenological introduction is followed by the discussion of the circular photogalvanic effect, the direct and inverse spin-galvanic effects and the trembling motion of spin-polarized electrons. Then we consider the pure spin currents and magneto-gyrotropic photocurrents. Finally, we discuss the spin-dependent photocurrents in topological insulators and Weyl semimetals.
\end{abstract}

\maketitle

\section{Introduction. Phenomenological Description} \label{ch7phenomdescrip}

The spin of electrons and holes in solid state systems is an
intensively studied quantum mechanical property showing a large
variety of interesting physical phenomena. One of the most
frequently used and powerful methods of generation and
investigation of spin polarization is optical
orientation~\cite{Meier}. Besides purely optical phenomena like
circularly-polarized photoluminescence, the optical generation of
an unbalanced spin distribution in a semiconductor may lead to
spin photocurrents.

Light propagating through a semiconductor and acting upon mobile
carriers can generate a $dc$ electric current, under short-circuit
condition, or a voltage, in the case of open-circuit samples. In
this review we consider only the photogalvanic effects
(PGE) which, by definition,
appear neither due to inhomogeneity of optical excitation of
electron-hole pairs nor due to inhomogeneity of the sample.
Moreover, we focus the attention here on the spin-photogalvanics
and discuss spin-related mechanisms of the following effects: the
circular PGE, spin-galvanic effect, inverse spin-galvanic effect
or spin polarization by electric current, generation of pure spin
photocurrents and magneto-gyrotropic photogalvanic effect.

The macroscopic features of all spin-dependent PGEs discussed in
this article, e.g., the possibility to generate a
helicity-dependent current, its behavior upon variation of
radiation helicity, crystallographic orientation, experimental
geometry etc., can be described in the frame of a phenomenological
theory which operates with conventional vectors, or {\it polar}
vectors, and pseudo-vectors, or {\it axial} vectors describing
rotation, and does not depend on details of microscopic mechanisms.
Below we consider one by one the phenomenological theory of
various spin-photogalvanic effects.

\paragraph{Circular Photogalvanic Effect}
The signature of photocurrent due to the circular PGE is that it
appears only under illumination with circularly polarized light
and reverses its direction when the sign of circular polarization
is changed. Physically, the circular PGE 
can be considered as a transformation of
the photon angular momenta into a translational motion of free
charge carriers. It is an electronic analog of mechanical systems
which transmit rotatory motion to linear one. In general there
exist two different possibilities for such transmission. The first
is based on the wheel effect: when a wheel being in mechanical
contact with the plane surface rotates it simultaneously moves as
a whole along the surface. The second transformation is based on
the screw effect and exemplified by a screw thread or a propeller.
In the both cases the rotation inversion results in the reversal
of motion, like the circular photogalvanic current
changes its sign following the inversion of the photon helicity
described by the degree of circular polarization
$P_c = \left( I_{\sigma_+} - I_{\sigma_-}\right) / \left( I_{\sigma_+} + I_{\sigma_-}\right)$, where $I_{\sigma_+}$ and $I_{\sigma_-}$ are the
intensities of the right-handed ($\sigma_+$) and left-handed ($\sigma_-$) circularly polarized radiation, respectively.

Phenomenologically, the circular photogalvanic current ${\bm j}$
is described by a second-order pseudotensor
\begin{equation} \label{pge}
j_{\lambda}
= I  \gamma_{\lambda \mu}  i\:
({\bm e} \times {\bm e}^*)_{\mu} = I P_c  \gamma_{\lambda \mu} n_{\mu}\:,
\end{equation}
where ${\bm n}$ is the unit vector pointing in the direction of
the exciting beam, $i$ is the imaginary
unity, $I$ and ${\bm e}$ are the light intensity and
polarization unit vector, for an elliptically polarized light the vector ${\bm e}$ is complex. In Eq.~(\ref{pge}), the second equality 
is valid for the transverse electromagnetic wave satisfying the property $i
({\bm e} \times {\bm e}^*) = P_c {\bm n}$.
Hereafter a repeated subscript is understood to imply summation over the range of this
subscript. In a bulk semiconductor or superlattice the index
$\lambda$ runs over all three Cartesian coordinates $x, y, z$. In
quantum well (QW) structures the free-carrier motion along the
growth direction is quantized and the index $\lambda$ enumerates
two in-plane coordinates. In quantum wires the free movement is
allowed only along one axis, the principal axis of the structure,
and the coordinate $\lambda$ is parallel to this axis. On the
other hand, the light polarization unit vector ${\bm e}$ and
directional unit vector ${\bm n}$ can be arbitrarily oriented in
space and, therefore, $\mu = x,y,z$. 

The tensor $\bm{\gamma}$
in~Eq.~(\ref{pge}) relates components of the polar vector ${\bm
j}$ and the axial vector ${\bm e} \times {\bm e}^*$ which means
that it is nonzero for point groups that allow optical activity or
{\it gyrotropy}. We remind that the gyrotropic point group
symmetry makes no difference between components of polar vectors,
like current or electron momentum, and axial vectors, like a
magnetic field or spin. Among 21 crystal classes lacking inversion
symmetry, 18 are gyrotropic; three nongyrotropic
noncentrosymmetric classes are T$_d$, C$_{3h}$ and D$_{3h}$, for more details see e.g.~\cite{Ivchenkobook2,GanTrushSchl}.

\paragraph{Spin-Galvanic and Inverse Spin-Galvanic Effects}

Another root of spin-photogalvanics is provided by optical spin
orientation. A uniform
nonequilibrium spin polarization obtained by any means, including
optical, yields an electric current if the system is characterized
by the gyrotropic symmetry. The current ${\bm j}$ and spin
{\boldmath$S$} are also related by a second-order pseudotensor
\begin{equation}
j_{\lambda} =
Q_{\lambda \mu} S_{\mu} \,\,\, .
\label{sge}
\end{equation}
This equation shows that the direction of the electric current is
coupled to the orientation of the nonequilibrium spin which is
given by the radiation helicity. The effect inverse to the
spin-galvanic effect is an electron spin polarization induced by a
$dc$ electric current, namely,
\begin{equation} S_{\mu} =
R_{\mu \lambda} j_{\lambda} \,\, .
\label{isge}
\end{equation}
We note the similarity of
Eqs.~(\ref{pge})-(\ref{isge}) characteristic for effects due to
gyrotropy: all three equations linearly couple a polar vector with
an  axial vector.

\paragraph{Pure Spin Photocurrents}

While describing spin-dependent phenomena, one needs, in addition
to electric currents, introduce spin currents.
Hereafter 
we use the
notation $q_{\lambda \mu}$ for the spin flux density with $\mu$
indicating the spin orientation and $\lambda$ indicating the flow
direction. Of special interest is the generation of pure spin
currents in which case a charge current
is absent but at least one of the $q_{\lambda \mu}$ components is
nonzero. A fourth-order tensor ${\bm P}$ relating the
pseudotensor ${\bm q}$ with the light intensity and polarization
as follows
\begin{equation} q_{\lambda \mu} = I P_{\lambda \mu \nu \eta} e_{\nu} e^*_{\eta}
\label{psc}
\end{equation}
has nonzero components in all systems lacking a center of
inversion symmetry. However an equivalence between polar and axial
vector components in the gyrotropic point groups suggests new
important mechanisms of pure spin currents connected with the
spin-orbit splitting of electronic bands.

\paragraph{Magneto-Photogalvanic Effects}
The variety of effects under consideration is completed by
magnetic-field induced photocurrents gathered in the class of
magneto-photogalvanic 
effects represented by the phenomenological
equation
\begin{equation} j_{\lambda} = I
\Phi_{\lambda \mu \nu \eta} B_{\mu} e_{\nu} e^*_{\eta} \,\,\, ,
\label{mge}
\end{equation}
where ${\bm B}$ is an external magnetic field. The symmetry
properties of the tensor ${\bm \Phi}$ coincide with those of ${\bm
P}$. In this paragraph we will consider a magneto-gyrotropic
photocurrent induced by a linearly polarized radiation which can
be directly connected with the pure spin current generated at zero
magnetic field.

\section{Circular Photogalvanic Effect} \label{ch7cpge}
\subsection{Historical Background}
The circular photogalvanic effect was independently
predicted by Ivchenko and Pikus \cite{Ivchenko78p640} and
Belinicher \cite{Belinicher}. It was first observed and studied in
tellurium crystals by Asnin et al.~\cite{Asnin}, see more
references in the book~\cite{sturman}.
In tellurium the current arises due to spin splitting  of the
valence band edge at the boundary of the first Brillouin-zone
(``camel back''structure). While neither bulk
zinc-blende materials like GaAs and related
compounds nor bulk diamond crystals like Si and Ge
allow this effect, in QW structures the circular PGE is possible
due to a reduction of symmetry. The circular PGE in gyrotropic QWs
was observed by Ganichev et al. applying terahertz (THz)
radiation~\cite{APL2000,PRL01,Ganichev03p935}.
In this review we discuss the circular PGE in QW structures grown
along the [001], [113] and [110] directions, present experimental
data for demonstration and outline the microscopic theory of the
effect under intersubband and interband optical transitions.

\begin{figure}[t]
\centering
\includegraphics*[width=6.7cm]{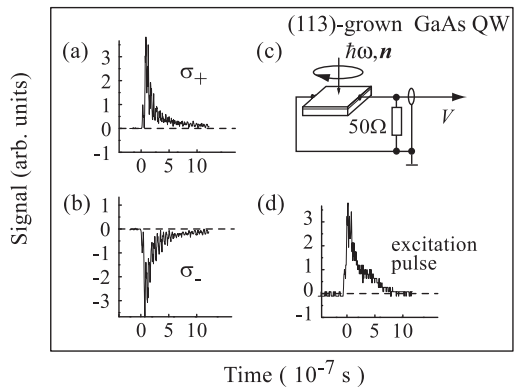}
\caption[]
{Oscilloscope traces obtained for pulsed excitation of (113)-grown $n$-type GaAs QWs
at $\lambda$ = 76 $\mu$m. ({\bf a}) and ({\bf b}) show circular PGE signals obtained for circular $\sigma_+$ and $\sigma_-$ polarization, respectively. For comparison, in (d) a signal pulse for a fast photon drag detector is plotted. In (c) the measurement arrangement is sketched.
After \protect\cite{Ganichev03p935}.}
\label{figure01_oscisignals}
\end{figure}

\begin{figure*}[t]
\centering
\includegraphics*[width=10.8cm]{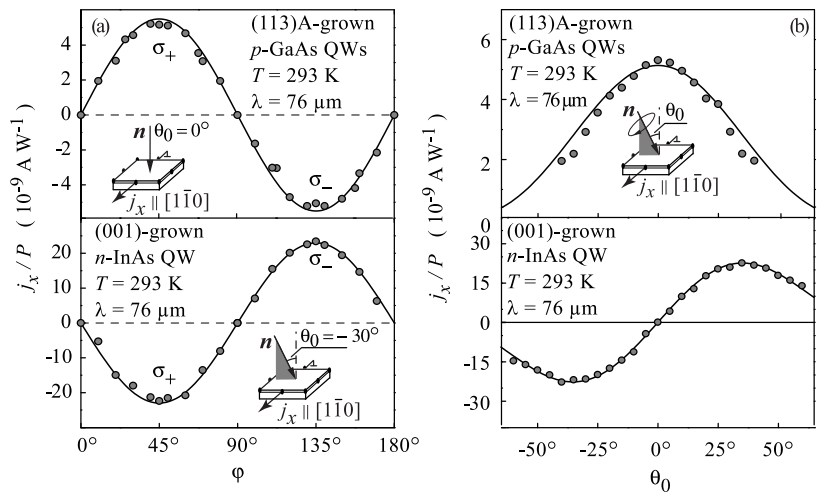}
\caption[]
{(a) Photocurrent in 
{
%\color{blue}
QWs normalized by the light power $P$ as
a function of the phase angle $\varphi$ defining the helicity}. {\it Upper
panel}: normal incidence of radiation on $p$-type $(113)$-grown
GaAs/AlGaAs QWs (symmetry class C$_s$). The current $j_x$ flows
along the $[1\bar{1}0]$ direction perpendicular to the mirror
plane. {\it Lower panel}: oblique incidence of radiation with an
angle of incidence $\theta_0 = - 30^{\circ}$ on $n$-type
(001)-grown InAs/AlGaSb QWs (symmetry class C$_{2v}$).
Full lines are fitted using one parameter according to~Eq.~(\protect \ref{pcphia}). (b) Photocurrent as
a function of the incidence angle $\theta_0$ for right-circularly
polarized radiation $\sigma_+$ measured perpendicularly to light
propagation.
{\it Upper panel}:
$p$-type
$(113)A$-grown GaAs/AlGaAs QWs.
{\it Lower panel}: $n$-type (001)-grown InAs/AlGaSb QWs.
Full lines represent theoretical fit. {
%\color{blue} 
The insets show the geometry of the experiment, ${\bm n}$ is the directional unit vector.}
After \protect\cite{PRL01}.}
\label{figure02_CPGE_2sym}
\end{figure*}

\subsection{Basic Experiments} \label{ch7basicexper}
With illumination of QW structures by polarized radiation a
current signal proportional to the helicity $P_c$ is generated in
unbiased samples. The irradiated structure represents a current
source wherein the current flows in the QW, see a scheme in
Fig.~\ref{figure01_oscisignals}(c).
Figures~\ref{figure01_oscisignals}(a)
and~\ref{figure01_oscisignals}(b) show measurements of the voltage
drop across a 50 Ohm load resistor in response to 100 ns laser
pulses at $\lambda$ = 76 $\mu$m. Signal traces are plotted for
right-handed (a) and left-handed circular polarization (b), in
comparison to a reference signal shown in
Fig.~\ref{figure01_oscisignals}(d) and obtained from a fast photon
drag detector~\cite{Ganichev03p935,GanichevPrettl}. The width  of the current
pulses is about 100~ns which corresponds to the THz laser pulses
duration.

Figure~\ref{figure02_CPGE_2sym}(a) presents results of measurements
carried out at room temperature on (113)-grown $p$-GaAs/AlGaAs
multiple QWs under normal incidence and (001)-grown
$n$-InAs/AlGaSb single QW structure under oblique incidence.
Optical excitation was performed by a high-power THz pulsed NH$_3$
laser operating at wavelength $\lambda = 76~\mu m$. The linearly
polarized light emitted by the laser could be modified to an
elliptically polarized radiation by applying a $\lambda/4$ plate
and changing the angle $\varphi$ between the optical axis of the
plate and the polarization plane of the laser radiation. Thus the
helicity $P_c$ of the incident light varies from $-1$ (left
handed, $\sigma_-$) to $+1$ (right handed, $\sigma_+$) according
to
\begin{equation} \label{pcphi}
P_c = \sin{2 \varphi}\:.
\end{equation}
One can see from Fig.~\ref{figure02_CPGE_2sym}(a) that the
photocurrent direction is reversed when the polarization switches
from right-handed circular, $\varphi = 45^{\circ}$, to
left-handed, $\varphi = 135^{\circ}$. The experimental points are
well fitted by the equation
\begin{equation} \label{pcphia}
j_{\lambda}(\varphi) =j_{\rm C} \sin{2 \varphi}
\end{equation}
with one scaling parameter. 
{
%\color{blue}
While in some systems and experimental conditions
the total photocurrent originates solely from the circular PGE, like for the
data in Fig.~\ref{figure02_CPGE_2sym}(a), in most cases it is accompanied by the
contributions caused by the linear photogalvanic or photon drag effects \cite{GanichevPrettl,Ivchenkobook2}. The latter are out of scope of this review focused on the spin photogalvanics. The additional contributions, however, complicate the dependence of the photocurrent on the radiation polarization state which in a
general case is given by
\begin{equation} \label{sincos}
j_{\lambda}(\varphi) = j_{\rm C}
\sin{2\varphi} + j_{\rm L1} \sin{4\varphi} + j_{\rm L2} \cos{4\varphi} + j_{\rm offset}\:.
\end{equation}
The circular photocurrent, being proportional to $P_c$, can easily
be extracted from the total current by subtracting the photocurrents excited by the
$\sigma_+$ and $\sigma_-$ circularly polarized radiation.}

In Fig.~\ref{figure02_CPGE_2sym}(b) closer look is taken at the
dependence of the photocurrent on the angle of incidence
$\theta_0$ in configuration with the incidence plane normal to the
axis $x \parallel [1 \bar{1} 0]$.

\begin{figure*}[t]
\centering
\includegraphics*[width=10.7cm]{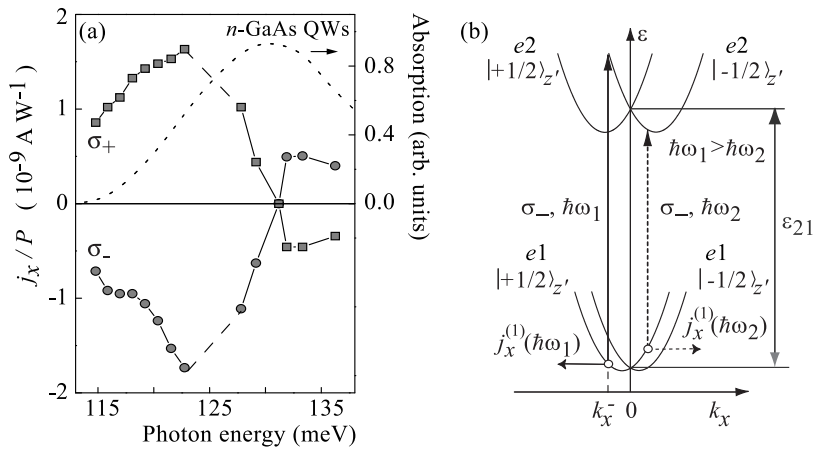}
\caption[]
{(a) Photocurrent in QWs normalized by the light power $P$
as a function of the photon energy $\hbar \omega$. Measurements are presented
for $n$-type (001)-grown GaAs/AlGaAs QWs of 8.2 nm width
%(symmetry class C$_{2v}$)
at room temperature. Oblique incidence of $\sigma_+$ (squares)
and $\sigma_-$ (circles) circularly polarized radiation with an angle of incidence
$\theta_0 =$ 20$^{\circ}$ was used. The current $j_x$ was measured perpendicular
to the light incidence plane $(y, z)$. The dotted line shows the
absorption measurement using a Fourier transform infrared spectrometer.
After \protect\cite{PRB03inv}.
(b) Microscopic picture describing the origin of the circular photogalvanic current and its spectral inversion
 in C$_s$ point group samples. The essential
ingredient is the splitting of the conduction band due to ${\bm k}$-linear
terms. Left handed circularly polarized radiation $\sigma_-$ induces direct
spin-flip transitions (vertical arrows) from the $e1$ subband with $s = 1/2$
to the $e2$ subband with $s'= - 1/2$. As a result an unbalanced
occupation of the $k_x$ states occurs yielding a spin polarized photocurrent.
For transitions with $k_x^-$ lying left to the minimum of $e1$ ($s
= 1/2$) subband the current indicated by $j_x$ is negative. At
smaller $\omega$ the transition occurs at a value of $k_x$ on the right-hand side of the subband
minimum, and the current reverses its sign. 
{
%\color{blue}
After~\cite{PRB03inv}
}.
}
\label{figure03_resonatCPGEinver}
\end{figure*}

For (113)-oriented QWs belonging to the symmetry class C$_s$ the
current retains its sign for all $\theta_0$ and
achieves its maximum at normal incidence, {
%\color{blue} 
see upper panel of
Fig.~\ref{figure02_CPGE_2sym}(b)}.  In contrast, in asymmetric
(001)-oriented samples (C$_{2v}$-symmetry) a variation of
$\theta_0$ in the plane of incidence normal to $x$ changes the
sign of the current $j_x$ for normal incidence, $\theta_0$=0, as
can be seen in the lower panel of Fig.~\ref{figure02_CPGE_2sym}(b).
Solid curves in this figure show a fit with
phenomenological equation~(\ref{pge}) adapted to a corresponding
symmetry and being in a good agreement with
experiment.

Further experiments demonstrate that circular PGE can be generated
by the radiation of wide range of frequencies from terahertz to
visible light. Applying light of various  frequencies the
photocurrent due to interband, intersubband and free carrier
absorption was detected.

Absorption of radiation in the range of 9 to 11 $\mu$m in $n$-type
GaAs QW samples of well widths 8$\div$9 nm is dominated by
resonant direct intersubband optical transitions
 between the first ($e1$) and the second
($e2$) size-quantized subbands.

Applying mid-infrared (MIR) radiation from the CO$_2$ laser, which causes direct transitions
in GaAs QWs, a current signal proportional to the helicity $P_c$
has been observed at normal incidence in (113)-oriented samples
and (110)-grown asymmetric QWs and at oblique incidence in
(001)-oriented samples, indicating a spin orientation induced
circular PGE~\cite{PRB03inv,JETP06}. In
Fig.~\ref{figure03_resonatCPGEinver}(a) the data are presented for a
(001)-grown $n$-type GaAs QW of 8.2 nm width measured at room
temperature. It is seen that the current for both left- and
right-handed circular polarization changes sign at the frequency
of the absorption peak. Spectral inversion of the photocurrent
direction at intersubband resonance has also been observed in
(113)-oriented samples and (110)-grown asymmetric
QWs~\cite{PRB03inv,JETP06}. {
%\color{blue} 

While in all above cases the geometry required
for the photocurrent is forced by symmetry arguments, in the systems of C$_1$ point-group
symmetry (containing only the identity operation), e.g. in (013)-oriented QWs, the space symmetry imposes no restriction
on the relation between radiation electric field and photocurrent components
resulting in a complex spectral and temperature behaviour \cite{C1}.}

\subsection{Microscopic Model for Intersubband Transitions}
\label{ch7microscmod}
{
%\color{blue}
We start the microscopic consideration from the intersubband mechanism of
circular PGE.} Microscopically, as shown below, a conversion of photon helicity
into a current as well  as a number of effects described in this
review is due to a removal of spin degeneracy in
the ${\bm k}$-space resulting in a shift of two
spin subbands as sketched in
Fig.~\ref{figure03_resonatCPGEinver}(b). Thus before a
discussion of the photocurrent origin we briefly describe the band
spin splitting.

\subsection{Relation to %$ {\bm k} $-Linear Terms
}

The linear in ${\bm k}$ terms in the Hamiltonian are given by the contribution
\begin{equation} \label{lkterms}
{\cal{H}}^{(1)}_{\bm k} = \beta_{\mu \lambda} \sigma_{\mu} k_{\lambda}
\end{equation}
to the expansion of the electron effective Hamiltonian in powers
of the wave vector ${\bm k}$. The coefficients $\beta_{\mu
\lambda}$ form a pseudotensor subjected to the same symmetry
restriction as the current-to-spin tensor ${\bm R}$ or the
transposed pseudotensors $\bm{\gamma}$ and ${\bm Q}$. The coupling
between the electron spin and momentum described by products of
the Pauli spin matrices $\sigma_{\mu}$ and the wave vector
components $k_{\lambda}$ as well as spin-dependent selection rules
for optical transitions yield a net
current sensitive to circularly polarized optical excitation. The first source of {\boldmath$k$}-linear terms is the bulk inversion asymmetry
(BIA), the corresponding contribution to  (\ref{lkterms}) is obtained by quantum-mechnical averaging \cite{Dyakonov86p110} of the cubic-in-${\bm k}$ Hamiltonian introduced by Dresselhaus~\cite{A1Dresselhaus55p580}. 
The spin-orbit splitting also arises from a structure inversion asymmetry   (SIA), see  
Refs.~\cite{Ganichev03p935,Vasko79,Bychkov84p78,Winkler03,Zawadzki2003pR1}. The breakdown of reduced symmetry at an interface, or the interface inversion asymmetry (IIA),  leads to a renormalization of the BIA contribution \cite{Krebs96p1829,roessler,Nest08}. 
The second-rank pseudotensors ${\bm \beta}, {\bm R}, {\bm \gamma}$ and ${\bm Q}$ for QW structures of different crystallographic orientations as well as application of the spin
photocurrents for studying the BIA and SIA anysotropy are reviewed in Ref. \cite{PhysStatSolGolub}.

\subsection{Circular PGE Due to Intersubband Transitions}

Figure~\ref{figure03_resonatCPGEinver}(b) illustrates the
intersubband transitions
$e1 \rightarrow e2$ resulting in the circular PGE. In order to
make the physics more transparent we will first consider the
intersubband circular photogalvanic current generated under normal
incidence in QWs of the $C_s$ symmetry, say, in (113)-grown QWs,
and use the relevant coordinate system $x \parallel [1\bar{1}0]$,
$y' \parallel [33\bar{2}]$, $z' \parallel [113]$. In the
linear-${\bm k}$ Hamiltonian we retain only the term $\sigma_{z'}
k_x$ because other terms make no contribution to the photocurrent
under the normal incidence. Therefore, the energy dispersion in
the $\nu$th electron subband depicted in
Fig~\ref{figure03_resonatCPGEinver}(b) is taken as
\begin{equation} \label{energynu}
E_{e \nu, {\bm k}, s}  = E^0_{\nu} + \frac{\hbar^2 k^2}{2 m_c} + 2 s
\beta_{\nu} k_x\:,
\end{equation}
where $s=\pm 1/2$ is the electron spin component along the
$z'$ axis, $\beta_{\nu} = \beta_{z'x}^{(\nu)}$ and,
for the sake of simplicity, we neglect nonparabolicity effects
assuming the effective mass $m_c$ to be the same in both subbands.

For the direct $e1$-$e2$ transitions shown in
Fig~\ref{figure03_resonatCPGEinver} by vertical arrows, the energy
and momentum conservation laws read
$$
E_{21} + 2 (s' \beta_2 - s \beta_1)k_x = \hbar \omega\:,
$$
where $E_{21}$ is the $\Gamma$-point gap $E^0_2 - E^0_1$ and $s',
s = \pm 1/2$.
As a result of optical selection rules the circular
polarization, e.g., left-handed, under normal incidence induces
direct optical transitions between the subband $e1$ with spin
$s=+1/2$ and the subband $e2$ with spin $s'=-1/2$. For
monochromatic radiation with photon energy $\hbar \omega_1 >
E_{21}$ optical transitions occur only at a fixed value of $k_x^-$
where the energy of the incident light matches the transition
energy as indicated by the arrow in
Fig.~\ref{figure03_resonatCPGEinver}(b). Therefore, optical
transitions induce an imbalance of the momentum distribution in
both subbands yielding an electric current along the $x$-direction
with the $e1$ and $e2$ contributions, antiparallel 
$({\bm j}^{(1)})$ or parallel (${\bm j}^{(2)}$) to
$x$, respectively. Since in $n$-type QWs the energy separation
between the $e1$ and $e2$ subbands is typically larger than the
energy of longitudinal optical phonons $\hbar\omega_{\rm LO}$, the
nonequilibrium distribution of electrons in the $e2$ subband
relaxes rapidly due to emission of phonons. As a result, the
electric current ${\bm j}^{(2)}$ vanishes and the current
magnitude and direction are determined by the group velocity and
the momentum relaxation time $\tau_p$ of uncompensated electrons
in the $e1$ subband with $s=+1/2$, i.e., by ${\bm j}^{(1)}$. By
switching circular polarization from left- to right-handed due to
selection rules light excites the spin down subband only. 

\begin{figure*}[t]
\centering
\includegraphics*[width=10.7cm]{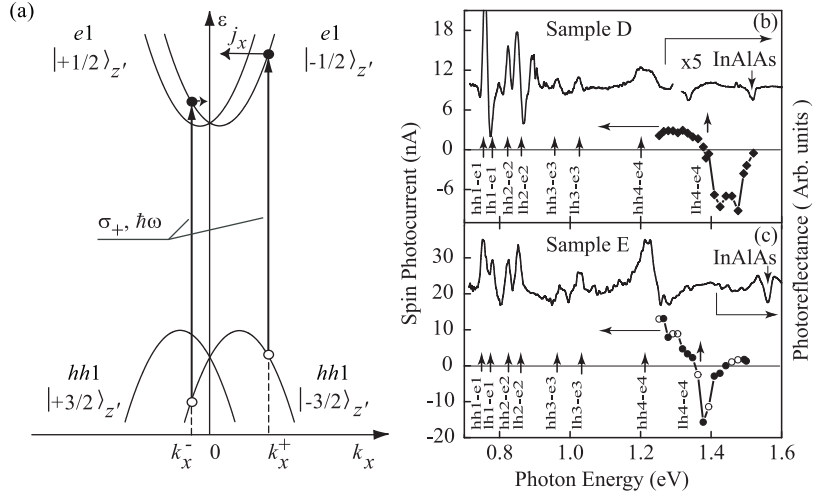}
\caption[]
{(a) Microscopic picture describing the origin
of interband circular PGE. The essential ingredient is the
spin splitting of the electron and/or hole states due to
linear-${\bm k}$ terms. (b)-(c) Spectral response of the circular
photogalvanic current observed in two (001)-grown In$_x$Ga$_{1-x}$As/InAlAs QW structures,
sample D and sample E, under oblique incidence. To enhance the
structure inversion asymmetry, sample E was grown with a graded indium composition
from 0.53 to 0.75 for the QW, instead of the
uniform indium composition of 0.70 for sample D. The
photoreflectance spectra of two samples are also shown to
determine the electronic structures of the samples. The arrows
indicate the heavy hole ($hh$) and light hole ($lh$) related
transitions. Data are given after \protect\cite{Yang06}.}
\label{figure04_interband}
\end{figure*}

Thus the whole picture mirrors and the current direction reverses. Spectral
inversion of the photocurrent at fixed helicity also follows from
the model picture of Fig.~\ref{figure03_resonatCPGEinver}(b). Indeed
decreasing the photon frequency to $\hbar\omega_2 < E_{21}$ shifts
the transitions toward positive $k_x$ and the direction of the
current reverses (horizontal dashed arrow).

Formally this process is described as follows. In the polarization
${\bm e} \perp z'$, the direct intersubband absorption is weakly
allowed only for the spin-flip transitions, $(e1, -1/2)
\rightarrow (e2, 1/2)$ for $\sigma_+$ photons and $(e1, 1/2)
\rightarrow (e2, - 1/2)$ for $\sigma_-$ photons. Particularly,
under the $\sigma_-$ photoexcitation the electrons involved in the
transitions have the fixed $x$-component of the wave vector
\begin{equation} \label{k21}
k_x^- = - \frac{\hbar \omega - E_{21}}{\beta_2 + \beta_1}
\end{equation}
and velocity
\begin{equation} \label{vx}
v^{(e \nu)}_x = \frac{\hbar k_x^-}{m_c} + (- 1)^{\nu + 1} \frac{\beta_{\nu}}{\hbar}\:.
\end{equation}
The circular photogalvanic current can be written as
\begin{equation} \label{jx}
j^{(e1)}_x = e \left( v^{(e2)}_x \tau_p^{(2)} - v^{(e1)}_x \tau_p^{(1)} \right)
\frac{\eta_{21} I}{\hbar \omega} P_c
\:,
\end{equation}
where $\tau^{(\nu)}_p$ is the electron momentum relaxation
time in the
$\nu$ subband, $\eta_{21}$ is the absorbance or the fraction of
the energy flux absorbed in the QW due to the transitions under
consideration, and minus in the right-hand side means that the
$e1$-electrons are removed in the optical transitions.

We assume that inhomogeneous broadening, $\delta_{21}$, of the resonance $E_{21}$ exceeds, by
far, the subband spin splitting. In this
case the convolution of the current given by Eq.~(\ref{jx}) with
the inhomogeneous
distribution function
leads to~\cite{{PRB03inv}}
\begin{eqnarray} \label{813}
j_x &=& \frac{e}{\hbar} (\beta_2 + \beta_1) \left[ \tau_p^{(2)}\:
\eta_{21}(\hbar \omega) \right. \\ &&\left. + (\tau_p^{(1)} - \tau_p^{(2)}) 
\: \langle E \rangle \: \frac{d
\:\eta_{21}(\hbar \omega)}{d\: \hbar \omega} \right] \frac{IP_c}{\hbar \omega}\:, \nonumber
\end{eqnarray}
where $\eta_{21}$ is the absorbance in the polarization ${\bm e}
\perp z'$ calculated neglecting the linear-${\bm k}$ terms but
taking into account the inhomogeneous broadening, $\langle E
\rangle$ is the mean value of the two-dimensional (2D) electron
energy, namely half of the Fermi energy $E_F$ for a degenerate 2D
electron gas and $k_B T$ for a nondegenerate gas.
Since the derivative $d \eta_{21}/d(\hbar \omega)$ changes its
sign at the absorption peak frequency and usually the time
$\tau^{(1)}_p$ is much longer than $\tau^{(2)}_p$,
the circular photogalvanic current given by~Eq.~(\ref{813}) exhibits the
sign-inversion behavior within the resonant
absorption
contour, in agreement with
the experimental observations~\cite{PRB03inv,JETP06}.

Similarly to the previously discussed case of the C$_s$ symmetry
the circular photogalvanic current for the $e1$-$e2$ transitions in
(001)-grown QWs exhibits the sign-inversion behavior within the
resonant absorption contour~\cite{{PRB03inv}}, in agreement with
the experimental data presented in
Fig.~\ref{figure03_resonatCPGEinver}(a).

\subsection{Interband Optical Transitions}
\label{interbandoptical}
For direct optical transitions between the heavy-hole valence subband
$hh1$ and conduction subband $e1$, the circular PGE is also most
easily conceivable in QWs of the $C_s$ symmetry which allows the
spin-orbit term $\beta_{z'x} \sigma_{z'} k_x$. 
{
%\color{blue}
We assume the parabolic dispersion in the conduction and heavy-hole subbands, $e1$  and $hh1$ respectively, taking into account the linear-${\bm k}$ terms as follows}
\begin{eqnarray} \label{betakx}
&&E_{e1, {\bm k}, \pm 1/2} = E_g^{QW} + \frac{\hbar^2 k^2}{2m_c} \pm
\beta_e k_x\:, \\ &&E^v_{hh1, {\bm k}, \pm 3/2} = - \frac{\hbar^2
k^2}{2m_v} \pm \beta_h k_x\:, \nonumber
\end{eqnarray}
where $m_v$ is the hole in-plane effective mass, $\beta_e =
\beta^{(e1)}_{z'x}$, $\beta_h = \beta^{(hh1)}_{z'x}$, $E_g^{QW}$
is the band gap renormalized because of the quantum confinement of
free carriers and the energy is referred to the valence-band top.
In Fig.~\ref{figure04_interband}(a) the allowed optical transitions
are from $j = - 3/2$ to $s=-1/2$ for the $\sigma_+$ polarization
and from $j = 3/2$ to $s=1/2$ for the $\sigma_-$ polarization.
Under circularly polarized radiation with a photon energy $\hbar
\omega$ and for a fixed value of $k_{y'}$ the energy and momentum
conservation allow transitions only from two values of $k_x$
labeled $k^-_x$ and $k^+_x$. The corresponding transitions are
shown in Fig.~\ref{figure04_interband}(a) by the solid vertical
arrows with their ``center-of-mass" shifted from the point
$k_x=0$. 

Thus the average electron velocity in the excited state
is nonzero and the contributions of $k^{\pm}_x$ photoelectrons to
the current do not cancel each other as in the case
$\beta_e=\beta_h=0$. Changing the photon helicity from $+1$ to
$-1$ inverts the current because the ``center-of-mass" for these
transitions is now shifted in the opposite
direction. The asymmetric distribution of photoelectrons in the
${\bm k}$-space decays within the momentum relaxation time.
However, under steady-state optical excitation new photocarriers
are generated resulting in a $dc$ photocurrent. The photohole
contribution is considered in a similar way. The final result for
the 
interband circular photogalvanic current can be presented
as
\begin{equation} \label{cpgeth}
j_x = - e ( \tau_p^{e} - \tau_p^{h} ) \left( \frac{\beta_e}{m_v} +
\frac{\beta_h}{m_c} \right) \frac{\mu_{cv}}{\hbar}  \frac{\eta_{eh}
I}{\hbar \omega} P_c\:,
\end{equation}
where $\eta_{eh}$ is the fraction of the photon energy flux
absorbed in the QW due to the $hh1 \rightarrow e1$ transitions and
$\tau_p^{e}, \tau_p^{h}$ are the electron and hole momentum
relaxation times. 

\begin{figure*}[t]
\centering
\includegraphics*[width=11.5cm]{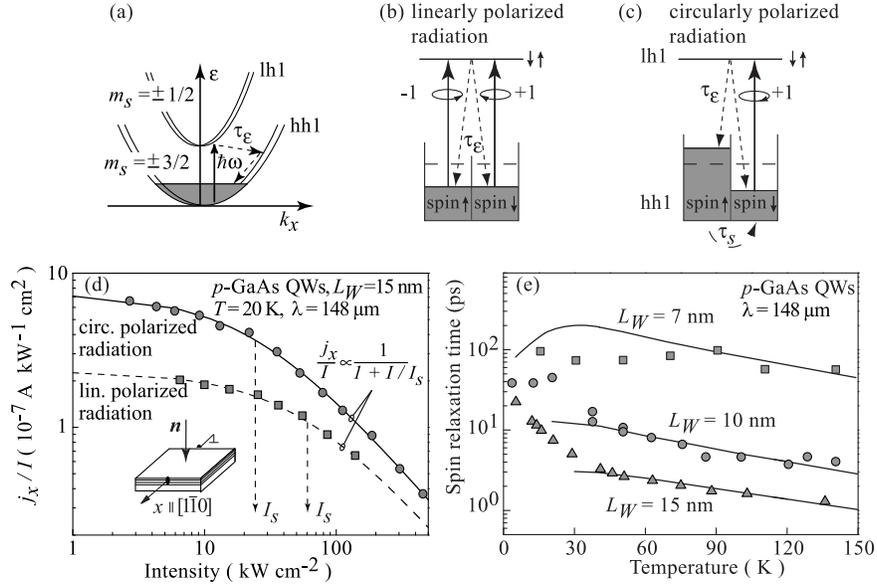}
\caption[]
{(a)-(c) Microscopic picture of spin sensitive
bleaching: Direct $hh1$-$lh1$ optical transitions (a) and
process of bleaching for two polarizations, linear (b) and circular (c).
Dashed arrows indicate energy ($\tau_\varepsilon$) and spin ($\tau_s$)
relaxation. (d) Circular (squares) and linear PGE (circles) currents $j_x$ normalized by the
intensity as a function of the intensity for circularly and
linearly polarized radiation. The inset shows the geometry of the experiment.
The measurements are fitted to $j_x/I \propto 1/(1 + I/I_s)$ with one parameter $I_s$
for each state of polarization. Data are given for (113)-grown samples
after \protect\cite{PRL02}.
(e) Spin relaxation times of holes for three different
widths of (113)-grown GaAs/AlGaAs QWs as a function of
temperature. The solid lines show a fit according to the
Dyakonov--Perel relaxation mechanism.
% The inset shows hole spin-splitting parameter b obtained from the fit.
After \protect\cite{Schneider04p420}.}
\label{figure05_spinsaturintensity}
\end{figure*}

The circular PGE at interband absorption was observed in GaAs-,
InAs- and GaN-based QW structures~\cite{Belkov283p2003,Bieler05,Yang06,Cho07,YuChen2011,YuChen2012,cpge2015} 
{
%\color{blue}
and bulk semiconductors InN \cite{cpgeInN} and BiTeBr \cite{giantRS}.}  In
Figs.~\ref{figure04_interband}(b) and~\ref{figure04_interband}(c)
diamonds and circles present the spectral dependence of the
circular photogalvanic current measured under interband optical transitions
between the higher valence and conduction subbands. The
photo-reflectance spectra of the samples are shown (solid curves)
to clearly indicate the quantized energy levels of electrons and
holes as marked by the arrows. For the both samples the
photocurrent spectral contour exhibits a change in sign, in a
qualitative agreement with the theoretical
prediction~\cite{Golub2003p235320}.

In addition to the interband and intersubband photocurrents, the circular PGE can be caused by  intraband (or intrasubband) mechanisms~\cite{PRL01}. In Refs.~\cite{SpivakArx,Orenstein}, a semiclassical theory of nonlinear transport and intraband photogalvanic effects has been proposed for noncentrosymmetric media, and the orbital Berry-phase contribution
to helicity-dependent photocurrents has been computed.

\subsection{Spin-Sensitive Bleaching} \label{ch7lincircdich}

Application of high intensities results in saturation (bleaching)
of PGE. This effect was observed for direct intersubband
transitions in $p$-type GaAs QWs and gave an
experimental access to spin relaxation
times~\cite{PRL02,Schneider04p420}. The method is based on the
difference in nonlinear behavior of circular and linear PGE. We remind that the
linear PGE is an another photogalvanic effect allowed in GaAs
structures and can be induced by linearly polarized
light~\cite{sturman,GanichevPrettl,Ivchenkobook2}. Both currents
are proportional to absorption and their nonlinear behavior
reflects the nonlinearity of absorbance.

Spin sensitive bleaching can be analyzed in terms of a simple model taking into
account both optical excitation and nonradiative relaxation
processes.
Excitation with THz radiation results in direct transitions
between heavy-hole ($hh1$) and light-hole ($lh1$) subbands,
{
%\color{blue} 
see Fig.~\ref{figure05_spinsaturintensity}(a)}. This process
depopulates and populates selectively spin states in the $hh1$ and
$lh1$ subbands. The absorption is proportional to the difference
of populations of the initial and final states. At high
intensities the absorption decreases since the photoexcitation
rate becomes comparable to the nonradiative relaxation rate to the
initial state.

% For figures use
\begin{figure*}[t]
\centering
\includegraphics*[width=10cm]{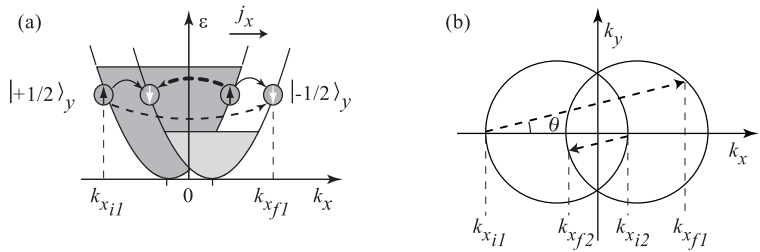}
\caption[]{Microscopic origin of the spin-galvanic current.
(a) One-dimensional sketch: the $\sigma_y k_x$ term
in the Hamiltonian splits the conduction band into two parabolas
with the spin $s_y = \pm 1/2$ pointing in the $y$-direction. If one
of the spin-split subbands is preferentially occupied, e.g., by spin injection
($|+1/2\rangle_y$-states in the figure), spin-flip
scattering results in a current in the $x$-direction. The scattering rate
depends on values of the initial and final electron wave vectors.  Thus,
the transitions sketched by dashed
arrows yield an asymmetric filling of the subbands and, hence, a
current flow. If instead of
the spin-up subband the spin-down subband is preferentially occupied,
the current direction is reversed. (b) The spin-flip transitions in the two dimensions
at scattering angle $\theta$ different from zero. 
{
%\color{blue}
After~\protect\cite{Ganichev03p935,Nature02}.} }
\label{figure06_modelsge}       % Give a unique label
\end{figure*}

Absorption of linearly polarized light is not spin
selective and the saturation is controlled by energy
relaxation,
{
%\color{blue} 
Fig.~\ref{figure05_spinsaturintensity}(b)}. In
contrast, absorption of circularly polarized light is spin
selective due to selection rules, and only one type of spin is
excited, Fig.~\ref{figure05_spinsaturintensity}(c). Note that during
energy relaxation the hot holes lose their photoinduced
orientation due to rapid relaxation so that the spin orientation
occurs only within the bottom of the $hh1$ subband. Thus the
absorption bleaching of circularly polarized radiation is governed
by energy relaxation of photoexcited carriers and spin relaxation
within the subband $hh1$. 
These processes are characterized by
energy and spin relaxation times,
$\tau_\varepsilon$ and
$\tau_s$, respectively.
If $\tau_s$ is longer than
$\tau_\varepsilon$ the bleaching of absorption becomes spin
sensitive and the saturation intensity of circularly polarized
radiation drops below the value of linear polarization.

Bleaching of absorption with
increasing the intensity of linearly polarized light is described
phenomenologically by the function
\begin{equation} \label{sat6}
\eta(I) = \frac{\eta_0}{1 + I/I_{se}}\:,
\end{equation}
where $\eta_0 = \eta(I \to 0)$ and $I_{se}$ is the characteristic
saturation intensity controlled by energy relaxation of the 2D
hole gas. Since the photocurrent of linear PGE, $j_{{\rm LPGE}}$, induced by the
linearly polarized light is proportional to $\eta I$, one has
\begin{equation} \label{sat7}
\frac{j_{{\rm LPGE}}}{I} \propto \frac{1}{1 + I/I_{se}}\:.
\end{equation}
The circular photogalvanic current $j_{{\rm CPGE}}$ induced by the circularly
polarized radiation is proportional to the degree of
hole spin polarization and given by~\cite{PRL02}
\begin{equation} \label{sat8}
\frac{j_{{\rm CPGE}}}{I} \propto \frac{1}{1 + I \left(I_{se}^{-1} + I_{ss}^{-1}\right)}\:,
\end{equation}
where $I_{ss} = p_s \hbar \omega/(\eta_0 \tau_s)$, $p_s$ is the 2D
hole density.

The measurements illustrated in
Fig.~\ref{figure05_spinsaturintensity}(d) indicate that the
photocurrent $j_x$ at a low power level depends linearly on the
light intensity and gradually saturates with increasing intensity,
$j_x \propto I/(1 + I/I_s)$, where $I_s$ is the saturation parameter.
One can see from Fig.~\ref{figure05_spinsaturintensity}(d) that the
measured saturation intensity for circular polarized radiation is
smaller than that for linearly polarized light. Using  the
measured values of $I_s$ and Eqs.~(\ref{sat7}) and~(\ref{sat8})
one can estimate the parameter $I_{ss}$ and even the time
$\tau_s$~\cite{PRL02,Schneider04p420}.

Figure~\ref{figure05_spinsaturintensity}(e) presents spin relaxation
times extracted from experiment
(points) together with a theoretical fit assuming that the
Dyakonov--Perel mechanism~\cite{DyakPerel} of 
hole spin relaxation is dominant.

\section{Spin-Galvanic Effect}
\label{ch9SGE}
The mechanisms of circular PGE discussed so far are linked with
the asymmetry in the momentum distribution of carriers excited in
optical transitions which are sensitive to the light circular
polarization due to selection rules. Now we discuss an additional
possibility to generate a photocurrent sensitive to the photon
helicity. In a system of free carriers with nonequilibrium
spin-state occupation but equilibrium energy distribution within
each spin branch, the spin relaxation
%or Larmor precession in an external magnetic field
can be accompanied by generation of an electric current.
This effect, predicted by Ivchenko et al.\cite{Ivchenko89p175},
was observed by Ganichev et al. applying THz radiation and named
the spin-galvanic 
effect~\cite{Nature02}.

If the nonequilibrium spin is produced by optical
orientation proportional to the
degree of light circular polarization $P_c$ the current generation
can be reputed just as another mechanism of the circular PGE.
However the nonequilibrium spin ${\bm S}$ can be achieved both by
optical and non-optical methods, e.g., by electrical spin
injection, and, in fact, Eq.~(\ref{sge}) presents an independent
effect. 

Usually the circular PGE and spin-galvanic effect are observed
simultaneously under illumination by circularly polarized light
and do not allow an easy experimental separation. However, they
can be separated in time-resolved measurements. Indeed, after
removal of light or under pulsed photoexcitation the circular
photogalvanic current decays within the momentum relaxation time
$\tau_p$ whereas the spin-galvanic
current decays with the spin relaxation time.

Another method which, on the one hand, provides a
uniform distribution in spin subbands and, on the other hand,
excludes the circular PGE was proposed in Ref.~\cite{Nature02}. It
is based on the use of optical excitation and the assistance of an
external magnetic field to achieve an in-plane polarization in
(001)-grown low-dimensional structures. 
{
%\color{blue} 
Finally we note that the spin-galvanic effect and its inversion, see Sect. \ref{inverse}, have been detected in electron transport experiments
which do not imply optical excitation and, therefore, are out of scope
of this review.

\subsection{Microscopic Mechanisms} \label{ch7microsmech}
For (001)-grown asymmetric QWs characterized by the C$_{2v}$
symmetry only two linearly independent components, $Q_{xy}$ and
$Q_{yx}$, of the tensor ${\bm Q}$ in Eq.~(\ref{sge}) are nonzero
so that
\begin{equation}
j_{x} = Q_{xy}S_y  \:,\: \: \: \: \: \: \: \: \: \: \: j_{y} =
Q_{yx}S_x\:. \label{sgec2v}
\end{equation}
Hence, a spin polarization driven current needs a spin component
lying in the QW plane.  For the C$_s$ symmetry of $(hhl)$-oriented
QWs, particularly, (113) and asymmetric (110), an additional
tensor component $Q_{xz^\prime}$ is nonzero and the spin-galvanic
current may be caused by nonequilibrium spins
oriented normally to the QW plane.

Figure~\ref{figure06_modelsge} illustrates the generation of a
spin-galvanic current. As already addressed above, it arises due
to {\boldmath$k$}-linear terms in the electron effective
Hamiltonian, see~Eq.~(\ref{lkterms}). For a 2D electron gas
system, these terms lead to the situation sketched in
Fig.~\ref{figure06_modelsge}(a). More strictly, the scattering
changes both $k_x$ and $k_y$ components of the electron wave
vector as shown Fig.~\ref{figure06_modelsge}(b) by dashed lines.
However, the one-dimensional sketch in \ref{figure06_modelsge}(a)
conveys the interpretation in a simpler and clearer way. In the
figure the electron energy spectrum along $k_x$ with allowance for
the spin-dependent term $\beta_{yx}\sigma_y k_x$ is shown. In this
case $s_y = \pm 1/2$ is a good quantum number. The electron energy
band splits into two subbands which are shifted in the
{\boldmath$k$}--space, and each of the bands comprises states with
spin up or down. Spin orientation in the $y$-direction causes the
unbalanced population in spin-down and spin-up subbands. As long
as the carrier distribution in each subband is symmetric around
the subband minimum point no current flows.

As illustrated in Fig.~\ref{figure06_modelsge}(a) the current flow
is caused by {\boldmath$k$}-dependent spin-flip relaxation
processes. Spins oriented in the $y$-direction are scattered along
$k_x$ from the more populated spin subband, e.g., the subband
$|+1/2 \rangle_y$, to the less populated subband $|-1/2
\rangle_y$. Four different spin-flip scattering events are
sketched in Fig.~\ref{figure06_modelsge}(a) by bent arrows. The
spin-flip scattering rate depends on the values of wave vectors of
the initial and final states\cite{Averkiev02pR271}.
Therefore, the spin-flip transitions shown by solid arrows in
Fig.~\ref{figure06_modelsge}(a) have the same rates. They preserve
the symmetric distribution of carriers in the subbands and, thus,
do not yield a current. However, two other scattering processes
shown by broken arrows are inequivalent and generate an asymmetric
carrier distribution around the subband minima in both subbands.
This asymmetric population results in a current flow along the
$x$-direction. 

The occurrence of a current is due to the spin
dependence of the electron scattering matrix elements $\hat{M}_{
\mbox{\footnotesize{\boldmath$k$}}^\prime
\mbox{\footnotesize{\boldmath$k$}}} = A_{
\mbox{\footnotesize{\boldmath$k$}}^\prime
\mbox{\footnotesize{\boldmath$k$}}  } \hat{I} + \mbox{\boldmath$
\sigma$} \cdot \mbox{\boldmath$ B$}_{
\mbox{\footnotesize{\boldmath$k$}}^\prime
\mbox{\footnotesize{\boldmath$k$}} } \:,$ where $A^*_{
\mbox{\footnotesize{\boldmath$k$}}^\prime
\mbox{\footnotesize{\boldmath$k$}} } =A_{
\mbox{\footnotesize{\boldmath$k$}}
\mbox{\footnotesize{\boldmath$k$}}^\prime}$, $B^*_{
\mbox{\footnotesize{\boldmath$k$}}^\prime
\mbox{\footnotesize{\boldmath$k$}}  } = B_{
\mbox{\footnotesize{\boldmath$k$}}
\mbox{\footnotesize{\boldmath$k$}}^\prime}$ due to hermicity of
the interaction and $A_{-
\mbox{\footnotesize{\boldmath$k$}}^\prime, -
\mbox{\footnotesize{\boldmath$k$}}  } =A_{
\mbox{\footnotesize{\boldmath$k$}}
\mbox{\footnotesize{\boldmath$k$}}^\prime}$, $B_{-
\mbox{\footnotesize{\boldmath$k$}}^\prime, -
\mbox{\footnotesize{\boldmath$k$}}  } = - B_{
\mbox{\footnotesize{\boldmath$k$}}
\mbox{\footnotesize{\boldmath$k$}}^\prime}$ due to the symmetry
under time inversion.
Within the model of elastic scattering the
current is not spin polarized since the same number of spin-up and
spin-down electrons move in the same direction with the same
velocity.
The spin-galvanic current can be estimated by~\cite{PRB03sge}
\begin{eqnarray}\label{kinetic}
j_x
& = & Q_{xy}S_y \sim e\: n_s
\frac{\beta_{yx}}{\hbar} \frac{\tau_p }{\tau^\prime_s}
S_y\:,
\end{eqnarray}
and the similar equation for $j_y$, where $n_s$ is the 2D electron
density, $\tau^\prime_s$ is the spin relaxation time due to the
Elliott--Yafet mechanism~\cite{Meier,Elliot,Yafet}.  Since spin-flip scattering is the origin
of the current given by Eq.~(\ref{kinetic}), this
equation is valid even if the Dyakonov--Perel
mechanism~\cite{Meier,DyakPerel} of spin relaxation dominates. The
Elliott--Yafet relaxation time $\tau^\prime_s$ is proportional to
the momentum relaxation time $\tau_p$. Therefore the ratio $\tau_p
/ \tau_s^\prime$ in Eq.\,(\ref{kinetic}) does not
depend on the momentum relaxation time. The in-plane average spin,
e.g., $S_y$ in Eq.\,(\ref{kinetic}), decays with the total spin
relaxation time $\tau_s$ and, hence, the time decay of the
spin-galvanic current following pulsed photoexcitation is
described by the exponential function $\exp{(- t/ \tau_s)}$. In
contrast, the circular PGE current induced by a short-pulse decays
within the momentum relaxation time $\tau_p$ allowing to
distinguish these two effects in time resolved measurements.

In general, in addition to the kinetic contribution to the current
there exists the so-called relaxational contribution which arises
due to the linear-${\bm k}$ terms neglecting the Elliott--Yafet
spin relaxation, i.e., with allowance for the Dyakonov--Perel
mechanism only. This contribution has the form
\begin{equation} \label{jintra}
{\bm j}
= - e n_s \tau_p \nabla_{\bm k} \left(\bm{
\Omega}^{(1)}_{\bm k}  \dot{{\bm S}} \right)  \:,
\end{equation}
where the spin rotation frequency $\bm{ \Omega}^{(1)}_{\bm k}$ is
defined by ${\cal{H}}^{(1)} = (\hbar/2) {\bm \sigma} \bm{
\Omega}^{(1)}_{\bm k}$, i.e., $\hbar \Omega^{(1)}_{{\bm k}, \mu} =
2 \beta_{\mu \lambda} k_{\lambda}$, and $\dot{{\bm S}}$ is the
spin generation 
{
%\color{blue}
(or production)} rate.

For optical transitions excited under oblique incidence of the
light in $n$-type zinc-blende-lattice QWs of the C$_{2v}$
symmetry, the spin-galvanic effect coexists with the circular PGE
described in Section~\ref{ch7cpge}. In the case of
intersubband transitions in (001)-grown QWs, the spin orientation
is generated by resonant spin-dependent and spin-conserving
photoexcitation followed by energy relaxation of the
photoelectrons from the subband $e2$ to $e1$ and their further
thermalization within the subband $e1$. The resulting spin
generation rate is given by a product of the optical transition
rate times the factor of depolarization, $\xi$, of the
thermalizing electrons, and the current $j_x$ is
estimated as
\begin{eqnarray}\label{intersge}
j_x
&& \sim
e\:\frac{\beta_{yx}}{\hbar} \frac{\tau_p
\tau_s}{\tau^\prime_s} \frac{\eta_{21}I}{\hbar \omega}
 P_c \xi n_y \:,
\end{eqnarray}
where $\eta_{21}$ is the absorbance under the direct transitions
$e1 \to e2$. Equation\,(\ref{intersge}) shows that the
spin-galvanic current is proportional to the absorbance and
determined by the spin splitting
constant in the first subband, $\beta_{yx}$ or
$\beta_{xy}$. This is in contrast to the circular PGE which is
proportional to the absorbance derivative, see Eq.\,(\ref{813}).

Finally we note that besides spin-flip mechanisms of the current generation
the spin-galvanic effect can be caused by 
the interference of spin-preserving
scattering and spin relaxation processes 
in a system of spin-polarized 
two-dimensional carriers~\cite{GolubSGE}. 
{
%\color{blue}
Burkov et al. \cite{PhysRevB.70.155308.pdf} have incorporated the spin-galvanic effect into a set of equations that provide a description of coupled spin and charge diffusive transport in a two-dimensional electron gas with the SIA spin-orbit  interaction. }

\subsection{Spin-Galvanic Photocurrent Induced by the Hanle Effect}
\label{Ch7SGEoptmag}
The spin-galvanic effect can be investigated by pure optical spin
orientation due to absorption of circularly polarized radiation in
QWs. However, the irradiation of QWs with circularly polarized
light also results in the circular PGE, and an indivisible mixture
of both effects may be observed since phenomenologically they are
described by the tensors, ${\bm \gamma}$ and ${\bm Q}$, equivalent
from the symmetry point of view. Nevertheless, microscopically
these two effects are definitely inequivalent. Indeed, the
spin-galvanic effect is caused by asymmetric spin-flip scattering
of spin polarized carriers and determined by the spin relaxation
processes. If spin relaxation is absent the spin-galvanic current
vanishes. In contrast, the circular PGE is a result of selective
photoexcitation of carriers in the {\boldmath$k$}-space with
circularly polarized light due to optical selection rules, it is
independent of the spin relaxation if $\tau_s \gg \tau_p$.

% For figures use
\begin{figure}[t]
\centering
\includegraphics*[width=7.35cm]{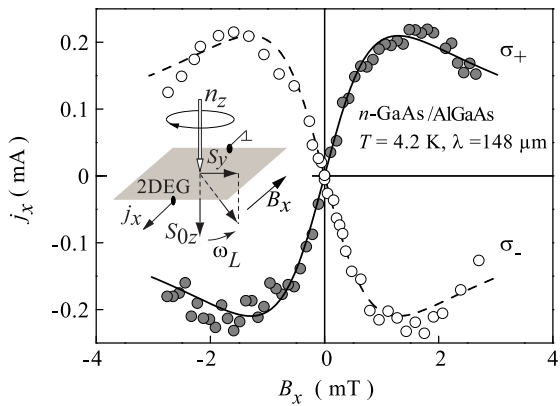}
\caption[]{Spin-galvanic current $j_x$ as a function of magnetic
field ${\bm B} \parallel x$ for normally incident right-handed (open circles) and
left-handed (solid circles) circularly polarized radiation.
Solid and dashed curves are fitted after
Eqs.\,(\protect \ref{sgec2v}) and\,(\protect
\ref{Hanle}) using the same value of the spin relaxation
time $\tau_s$ and scaling of the ordinate. Inset shows optical scheme
of generating a uniform in-plane spin polarization which causes a
spin-galvanic current. Electron spins
are oriented normal to the QW plane by circularly polarized
radiation and rotated into the plane by the Larmor precession in an
in-plane magnetic field $B_x$. After\protect\cite{Nature02}.}
\label{figure07_Hanle}       % Give a unique label
\end{figure}

Here we describe a method which, on the one hand, achieves a
uniform distribution of nonequilibrium spin polarization by
optical means and, on the other hand, excludes the circular
PGE~\cite{Nature02}. The polarization is obtained by absorption of
circularly polarized radiation at normal incidence on (001)-grown
QWs as depicted in the inset in Fig.~\ref{figure07_Hanle}. For
normal incidence the spin-galvanic effect as well as the circular PGE vanish
because both $S_x=S_y=0$ and $n_x=n_y=0$. Thus, the spin
orientation $S_{0z}$ along the $z$-axis is achieved but no
spin-induced photocurrent is generated,
{
%\color{blue} 
in difference with QWs of other 
crystallographic orientations, e.g. (110), (113) or (013).} 

An in-plane spin component, necessary for the spin-galvanic
effect 
{
%\color{blue}
in (001) oriented QWs}, arises in a magnetic field {\boldmath$B$}\,$\parallel x$. The field perpendicular to the initially oriented spins rotates
them into the plane of the 2D electron gas due to the Larmor
precession (Hanle effect). The nonequilibrium
spin polarization $S_y$ is given by
\begin{equation}
S_y = - \frac{\omega_L \tau_{s\perp} }{1 + (\omega_L \tau_s )^2}\:
S_{0z}\:,
\label{Hanle}
\end{equation}
where $\tau_s = \sqrt{\tau_{s \parallel} \tau_{s \perp} }$,
$\tau_{s\parallel}$ and $\tau_{s\perp}$ are the
longitudinal and transversal electron spin relaxation times, and
$\omega_L$ is the Larmor frequency. Since
in the experimental set-up $\dot{{\bm S}}$ is parallel to $z$, the
scalar product 
{
%\color{blue}
$(\bm{ \Omega}^{(1)}_{\bm k}\dot{{\bm S}})$} vanishes
and, according to Eq.~(\ref{jintra}), the spin-galvanic effect is
not contributed by the relaxational mechanism and arises only due
to the kinetic mechanism described by
Eq.~(\ref{kinetic}). The observation of the Hanle
effect, 
{
%\color{blue}
see Fig.~\ref{figure07_Hanle},} demonstrates that free carrier intrasubband
transitions can polarize
the spins of electron systems.
The measurements allow one to extract the spin relaxation time
$\tau_s$ from the peak position of the photocurrent reached at
$\omega_L \tau_s = 1$ providing experimental access to
investigation of spin relaxation times for monopolar spin
orientation,
 where only one type of charge carriers is involved in
the excitation--relaxation
process~\cite{Nature02,PASPS02monop}. This condition is close to that of
electrical spin injection in semiconductors.

We note that a similar set-up was applied in experiments of Bakun et
al.~\cite{Bakun84p1293} carried out on bulk AlGaAs excited by
interband absorption and demonstrating spin photocurrents  caused
by the inhomogeneous spin distribution predicted
in~\cite{Averkiev83p393,Dyakonov71p144}, known now as the 
Inverse Spin Hall Effect.  The crucial difference to the spin-galvanic effect is that in
the case of surface photocurrent caused by optical
orientation a gradient of spin density is needed. Naturally this
gradient is absent in QWs where the spin-galvanic effect has been
investigated because QWs are two-dimensional and have no
``thickness''.

% For figures use
\begin{figure*}[t]
\centering
\includegraphics*[width=10.5cm]{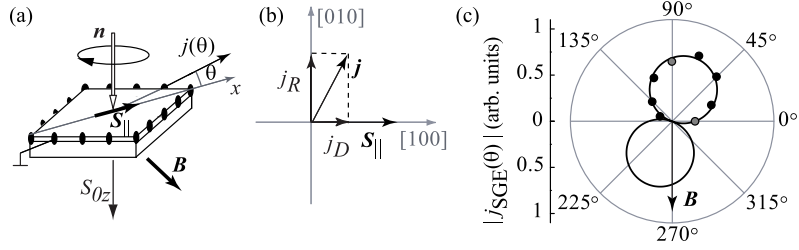}
\caption[]{The separation of the 
 SIA and BIA
contributions to the spin-galvanic effect observed in an $n$-type InAs single QW at room
temperature for the case of the electron spin ${\bm S}_{\parallel} \parallel [100]$.
(a) Geometry of the experiment. (b) The direction of  the
 SIA and BIA
contributions to the photocurrent. (c) The spin galvanic current measured as a function of the angle $\vartheta$
between the pair of contacts and the $x$ axis.
After~\protect\cite{PRL04}. }
\label{figure11_RD}       % Give a unique label
\end{figure*}

For optical excitation of the spin-galvanic effect mid-infrared,
far-infrared (terahertz frequencies) and visible laser radiation has
been used~\cite{Ganichev03p935,Ivchenkobook2,GanichevPrettl}. Most
of the measurements were carried out in the long wavelength range
with photon energies less than the energy gap of investigated
semiconductors. The advantage is that, in contrast to interband
excitation resulting in the valence band -- conduction band
transitions, there are no spurious photocurrents due to other
mechanisms like the Dember effect, photovoltaic effects at
contacts and Schottky barriers etc.

In contrast to the circular PGE the spin-galvanic effect induced
by the Hanle effect under intersubband transitions excited by the mid-infrared radiation
does not change its sign with frequency radiation and follows the
spectral behavior of direct intersubband
absorption~\cite{PASPS02monop}.
This result is in agreement with the mechanism of the
spin-galvanic effect discussed in the previous subsection, see
Eqs.~(\ref{intersge}), and clearly demonstrates that
this effect has different microscopic origin.
The observation of the mid-infrared and terahertz radiation excited spin-galvanic
effect, which is due to spin orientation, gives clear evidence
that direct intersubband and Drude absorption of circularly
polarized radiation result in a monopolar spin
orientation. Mechanisms
of the monopolar spin orientation were analyzed in
Refs.~\cite{PASPS02monop,Ivchenko04p379}. We would like to
emphasize that spin-sensitive $e1$-$e2$ intersubband transitions
in (001)-grown $n$-type QWs have been observed at normal incidence
when there is no component of the electric field of the radiation
normal to the plane of the QWs.

\subsection{Spin-Galvanic Effect at Zero Magnetic Field}
\label{Ch7SGEopticpure}

In the experiments described above an external magnetic field was
used for re-orientation of an optically generated spin
polarization.  The spin-galvanic effect can also be generated at
optical excitation only, without application of an external
magnetic field. The necessary in-plane component of the spin
polarization can be obtained by oblique incidence of the exciting
circularly polarized radiation but in this case the circular PGE
may also occur interfering with the spin-galvanic effect. However,
the spin-galvanic effect due to pure optical excitation was
demonstrated making use the difference in spectral behavior of
these two effects excited by inter-subband transitions in $n$-type
GaAs QWs~\cite{PRB03sge}. Experiments have been carried out
making use of the spectral tunability of the free electron laser
``FELIX''. The helicity
dependent photocurrent closely following the absorption spectrum
was detected demonstrating dominant contribution of
the spin-galvanic effect.

\subsection{Determination of the SIA/BIA Spin Splitting Ratio}
\label{ch7spincurrentappl}

An important application of the spin-galvanics was addressed
in~\cite{PRL04}. It was demonstrated that angular dependent
measurements of spin photocurrents allow one to separate the different contributions to the spin-orbit Hamiltonian (\ref{lkterms}). Later on this method 
was extended and improved by using measurements of both the spin-galvanic effect
(at normal incidence of the radiation in the presence of an in-plane magnetic field) and circular PGE
(at oblique incidence with no magnetic field applied) \cite{PRB07gig,PhysStatSolGolub}. 

Experiments were carried out on (001)-oriented QWs for which the
Hamiltonian Eq.~(\ref{lkterms}) for the first subband reduces to
\begin{equation} \label{biasia}
{\cal{H}}^{(1)}_{\bm k} = \alpha(\sigma_{x_0} k_{y_0} - \sigma_{y_0} k_{x_0}) +
\beta(\sigma_{x_0} k_{x_0} - \sigma_{y_0} k_{y_0})
 \:,
\end{equation}
where $x_0, y_0$ are the crystallographic axes $[100]$ and $[010]$,
the parameters $\alpha$ and $\beta$ result from the structure-inversion and bulk-inversion
asymmetries, SIA and BIA, respectively. The
%Following Refs.~\cite{Pikus1,Pikus2} the 
first and second contributions to the Hamiltonian (\ref{biasia}) are often called the Rashba and Dresselhaus terms.

Note that, in the coordinate system with
$x \parallel [1\bar{1}0]$ and $y\parallel [110]$, the matrix $
{\cal{H}}^{(1)}_{\bm k}$ gets the form $\beta_{xy} \sigma_x k_y +
\beta_{yx}  \sigma_y k_x$ with $\beta_{xy} = \beta + \alpha$,
$\beta_{yx} = \beta - \alpha$. According to Eq.~(\ref{kinetic})
the current components $j_x$, $j_y$ are proportional,
respectively, to $\beta_{xy}$ and $\beta_{yx}$
and, therefore, angular dependent measurements of spin
photocurrents allow one to separate the  SIA and BIA linear in-${\bm k}$
terms. By mapping the magnitude of the spin photocurrent in the QW
plane the ratio of both terms can directly be determined from
experiment and does not rely on theoretically obtained quantities.
The relation between the photocurrent and spin directions can be
conveniently expressed in the following matrix form
\begin{equation}
\label{bissiam} \mbox{\boldmath$j$}  \propto \left(
{{\begin{array}{*{20}c}
 \beta \hfill  & \,\,\, -\alpha \hfill \\
  \alpha  \hfill  & \,\,\, - \beta \hfill \\
\end{array} }} \right)\mbox{\boldmath$S_\parallel$} \:,
\end{equation}
where $\mbox{\boldmath$j$}$ and
$\mbox{\boldmath$S_\parallel$}$ are two-component columns with the
in-plane components along the crystallographic axes $x_0 \parallel
[100]$ and $y_0 \parallel [010]$. The directions of the
 SIA and BIA
coupling induced photocurrents are shown in
Fig.~\ref{figure11_RD}(b) for the particular case ${\bm
S}_{\parallel} \parallel [100]$.

Figure~\ref{figure11_RD}(c) shows the angular dependence of the
spin-galvanic current $j(\vartheta)$
measured on an $n$-type (001)-grown InAs/Al$_{0.3}$Ga$_{0.7}$Sb
single QW  of 15\,nm width at room temperature. Because of the
admixture of photon helicity-independent magneto-gyrotropic
effects (see Section~\ref{MGEf}) the spin-galvanic effect is
extracted after eliminating current contributions which are
helicity-independent: $j
=\left(j_{\sigma_+}-j_{\sigma_-}\right)/2$.

The sample edges are oriented along the $[1\bar{1}0]$ and [110]
crystallographic axes. Eight pairs of contacts on the sample allow
one to probe the photocurrent in different directions, see
Fig.~\ref{figure11_RD}(a). The optical spin orientation was
performed by using a pulsed molecular NH$_3$ laser. The
photocurrent ${\bm j}$ is measured in the unbiased
structure in a closed circuit configuration. The nonequilibrium
in-plane spin polarization {\boldmath$S_\parallel$} is prepared as
described in Section~\ref{Ch7SGEoptmag}, see also
Fig.\ref{figure11_RD}(a). The angle between the magnetic field and
{\boldmath$S_\parallel$} can in general depend on details of the
spin relaxation process. In these particular InAs QW structures
the isotropic Elliott--Yafet spin relaxation
mechanism
dominates. Thus, the in-plane spin polarization
{\boldmath$S_\parallel$} is always perpendicular to {\boldmath$B$}
and can be varied by rotating {\boldmath$B$} around $z$ as
illustrated in Fig.~\ref{figure11_RD}(a). The circle in
Fig.~\ref{figure11_RD}(c) represents the angular dependence
$\cos{(\vartheta - \vartheta_{\rm max})}$, where
$\vartheta$ is the angle between the pair of contacts
and the $x$ axis and $\vartheta_{\rm max} = \arctan{(j_R/j_D)}$, with $j_R$ and $j_D$ 
being the SIA and BIA contributions to the photocurrent, respectively.
The best fit in this sample is achieved for the ratio
$j_R/j_D=\alpha/\beta = 2.1$. The method was also used for
investigation of  SIA/BIA ${\bm k}$-linear spin-splitting in GaAs
heterostructures where spin relaxation  is controlled by
Dyakonov--Perel mechanism~\cite{PRB07gig}. These experiments
demonstrate that growth of structures with various delta-doping
layer position accompanied by experiments on spin-galvanic effect
makes possible a controllable variation of the structure inversion
asymmetry and preparation of
samples with equal  SIA and BIA
constants or with a zero SIA constant. 

The measurements of the spin-galvanic and circular photogalvanic
effects in a set of InGaAs/InAlAs QW structures with semitransparent
gate, supported by the weak antilocalization experiments, permitted
to find a proper QW design for the realization of the persistent helix condition $|\alpha| = |\beta|$
of the parameters in Eq.~(\ref{biasia}) \cite{helix}. Application of the two complementary experiments, transport and photogalvanic, enabled to extract information on the role of cubic-in-${\bm k}$ terms on spin transport in a material with strong spin-orbit interaction, for details see the review \cite{PhysStatSolGolub}.

\begin{figure*}[t]
\centering
\includegraphics*[width=12cm]{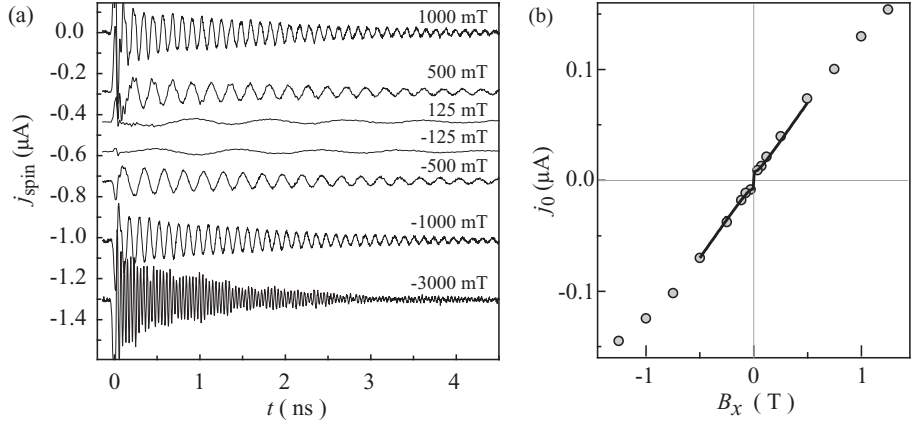}
\caption[]{
{
%\color{blue}
Magnetic field dependence of spin precession driven coherent Zitterbewegung. ({\bf a}) Time-resolved \textit{ac} current along the [110] direction in the InGaAs(001) epilayer at various magnetic fields. Larmor precession frequency and magnitude of \textit{ac} current increase with increasing magnetic field strength while the sign of the \textit{ac} current reverses when reversing the magnetic field direction. ({\bf b}) Amplitude $j_0$ of \textit{ac} current vs. applied magnetic field $B_x$. The solid line is a fit to Eq.~(\ref{amplit}).} 
{
%\color{blue} 
After \cite{zitter}.}
}
\label{fig2}       % Give a unique label
\end{figure*}

{
%\color{blue}

\subsection{Coherent Trembling Motion of Spin-Polarized Electrons}
Equation (\ref{jintra}) is derived for the steady-state spin generation. If the electron spin ${\bm S}$
varies in time then this equation reads
\begin{equation} \label{jintra2}
{\bm j}
= e n_s \tau_p \nabla_{\bm k} \left[\bm{
\Omega}^{(1)}_{\bm k} \left( - \dot{{\bm S}} + \frac{d {\bm S}(t)}{d t}\right) \right] \:.
\end{equation}
Particularly, under a short-pulse interband excitation at $t=0$ by circularly polarized light, the 
spin-galvanic
 current at $t > 0$ has the form
\begin{equation} \label{jintra3}
{\bm j} = e n_s \tau_p \nabla_{\bm k} \left(\bm{\Omega}^{(1)}_{\bm k}\frac{d {\bm S}(t)}{d t}\right)  \:.
\end{equation} 
In the normal-incidence geometry and in the presence of an in-plane magnetic field ${\bm B} \parallel x$ the electron spin ${\bm S}(t)$ processes around the field and exhibits an oscillatory behaviour 
\begin{eqnarray}
&&S_z(t) = S_z(0) {\rm e}^{- t/\tau_s} \cos{(\omega_L t)} \:,\\&&S_y(t) = - S_z(0) {\rm e}^{- t/\tau_s}\sin{(\omega_L t)}\:. \nonumber 
\end{eqnarray}
Substituting} $S_y(t)$ into Eq.~(\ref{jintra3}) we find for the $x$-component of the electric current
\begin{eqnarray} \label{jxs0}
j_x(t) &=& 2 e n_s \tau_p \frac{\beta_{yx}}{\hbar} S_z(0) {\rm e}^{- t/\tau_s}\\&& \times\left[ \frac{1}{\tau_s}\sin{(\omega_L t)} - \omega_L \cos{(\omega_L t)} \right] \:. \nonumber
\end{eqnarray}

Stepanov et al. \cite{zitter} have measured under identical experimental conditions the time-resolved spin-Faraday rotation
 of electron spin precession $S_z(t)$ and time-dependent {\it ac} difference current $j_{\rm spin} = [j(\sigma_+) - j(\sigma_-)]/2$ induced by the pump pulses with $\sigma_+/\sigma_-$ helicities. The samples consisted of a 500 nm thick In$_{0.07}$Ga$_{0.93}$As epilayer grown on a semi-insulating (001) GaAs wafer. Due to the strain caused by the lattice mismatch and the reduced symmetry of the structure, the spin-orbit terms in the electron effective Hamiltonian have the form similar to Eq.~(\ref{biasia}). A series of the current measurements at different values of the magnetic field are shown in Fig.~\ref{fig2}(a). The current can be fitted by an exponentially decaying cosine function
$$
j_x(t) = j_0  {\rm e}^{- t/\tau_s} \cos{(\omega_L t + \phi)}\:,
$$
with the amplitude 
\begin{equation} \label{amplit}
j_0 \propto \beta_{yx}S_z(0) {\rm sign}(B_x) \sqrt{\omega_L^2 + \tau_p^{-2}}
\end{equation}
and the phase $\phi = - \cot^{-1}{(\omega \tau_s)}$. Figure \ref{fig2}(b) presents the experimental and calculated dependences of the amplitude $j_0$ on the magnetic field.

The ac current (\ref{jxs0}) can be understood in terms of the coherent trembling motion
 (Zitterbewegung)
 of spin-polarized electrons~\cite{zitter}. The trembling motion originates from the fact that, in quantum mechanics, the electron velocity is not a conserved quantity in the presence of spin-orbit interaction. Taking into account the linear in ${\bm k}$ terms (\ref{lkterms}) and the Zeeman Hamiltonian $(\hbar/2) \omega_L \sigma_x$ we have for the electron velocity $v_x = (\hbar k_x/m_c) + (\beta_{yx}/\hbar) \sigma_y$ and then obtain for the acceleration
$$
\frac{d v_x}{d t} = -  \frac{\beta_{yx}}{\hbar} \omega_L \sigma_z\:.
$$
Including the electron momentum scattering as a ``friction force'' and neglecting the spin relaxation we arive at the following equation for the average electron velocity
$$
\frac{d v_x}{d t} + \frac{v_x}{\tau_p}= -  \frac{2 \beta_{yx}}{\hbar} \omega_L S_z\:.
$$
In the fast scattering limit, $\omega_L \tau_p \ll 1$, the oscillating current is given by
\begin{equation}
j_x = e n_s v_x(t) = -  2 n_s \tau_p \frac{\beta_{yx}}{\hbar} \omega_L S_z(0)\cos{(\omega_L t)}\:,
\end{equation}
in agreement with Eq.~(\ref{jxs0}) where $\tau_s$ is set to infinity. }

\section{Inverse Spin-Galvanic Effect} \label{inverse}
The effect inverse
to the spin-galvanic effect is the electron spin polarization generated
by a charge current ${\bm j}$.
First it was predicted in
\cite{Ivchenko78p640} and observed in bulk tellurium
\cite{Vorobjev79p441},
{
%\color{blue} 
see also the recent paper \cite{Tellurium2012}.
The symmetry of bulk tellurium belongs to the gyrotropic crystal class $D_{3}$ 
which contains the right or left handed threefold screw axis, respectively the space groups $D^4_3$ and $D^6_3$. The extrema of the conduction and valence are located at the vortices of the hexagonal Brillouin zone, the points $H$ and $H'$ ($M$ and $P$ in other notation). The degeneracy in the valence band is completely lifted, the two upper nondegenerate subbands $H_4$ and $H_5$ are split by $2 \Delta = 126$ meV. The hole wave function in the subband $H_4$ is a superposition of the two Bloch states $| \pm 3/2 \rangle$
\[
\Psi_{H_4, k_z} = C_{\frac32}(k_z)| 3/2 \rangle + C_{-\frac32}(k_z)| - 3/2 \rangle\:,
\]
where
\[
C_{\pm \frac32}(k_z) = \sqrt{\frac{E \pm \beta k_z}{2E}}\:,\: E = \sqrt{\Delta^2 +  (\beta k_z)^2}\:,
\]
and the spin-orbit coefficient $\beta = 2.4 \times 10^{-8}$ eV$\cdot$cm.
The average component of the angular momentum on the principal axis $z$ is given by
\begin{equation} \label{Szkz}
S_z(k_z) = \frac12 \left( C^2_{\frac32} - C^2_{-\frac32} \right) = \frac{\beta k_z}{2E}\:.
\end{equation}
An external cw electric field ${\bm F}$ applied in the $z$ direction shifts the hole distribution in the ${\bm k}$-space by the drift wave vector $\bar{k}_z \propto F_z$ and induces the free-carrier spin
$S_z \approx \beta \bar{k}_z/2 \Delta$. The appearance of the current-induced spin induces the rotation of the polarization plane of the linearly polarized probe light \cite{Vorobjev79p441,Tellurium2012} and the circular polarization of the interband photoluminescence.

In $p$-type tellurium, the energy splitting between the nondegenerate valence subbands $H_4$ and $H_5$  exceeds the uncertainty $\hbar/\tau_p$, the spin ${\bm S}$ and the wave vector ${\bm k}$ are strictly coupled, and the off-diagonal intersubband components of the hole density matrix are negligible. In the opposite limiting case where the splitting of the conduction band due to the linear ${
\bm k}$ terms (\ref{lkterms}) is small as compared to $\hbar/\tau_p$, the off-diagonal components of the electron spin density matrix $\rho_{\bm k}$ are not suppressed by the splitting and the relation between the nonequilibirum spin and the wave vector is more loose. The spin orientation by current in this limit of fast momentum scattering was theoretically demonstrated for QW systems by  Vas'ko and Prima~\cite{Vasko}, Levitov et al.~\cite{levitov}, Aronov and Lyanda-Geller~\cite{Aronov89p431} and Edelstein~\cite{Edelstein89p233}. The study was extended in
Refs.~\cite{Aronov91p537,Chaplik,VaskoRai,TarJETP,Schlieman07,Raichev,Tarasenko2008,GolIvch,Raimondi2016,Gorini2017}, see the review \cite{GanTrushSchl} for more details. The first} direct experimental proofs of this effect
were obtained in semiconductor
QWs~\cite{Ganichev04p0403641,Silov2004}
as well as in strained bulk material~\cite{Kato04p176601}.
At present inverse spin-galvanic effect has been observed in
various low-dimensional structures based on GaAs, InAs, ZnSe and
GaN.

Phenomenologically, the averaged nonequilibrium free-carrier spin
${\bm S}$ is linked to ${\bm j}$ by Eq.~(\ref{isge}).
Microscopically, the spin polarization can be found from the
kinetic equation for the electron spin density matrix $\rho_{\bm k}$ which can conveniently be presented
in the form
\begin{equation} \label{denmat}
\rho_{\bm k} = f_{\bm k} + {\bm s}_{\bm k} {\bm \sigma}\:,
\end{equation}
where $f_{\bm k} = {\rm Tr}\{ \rho_{\bm k}/2 \}$ is the
distribution function and ${\bm s}_{\bm k} = {\rm Tr}\{ \rho_{\bm
k} {\bm \sigma}/2 \}$ is the average spin in the ${\bm k}$ state.
In the presence of an electric field ${\bm F}$ the kinetic
equation reads
\begin{equation} \label{denkin}
\frac{e {\bm F}}{\hbar} \frac{\partial \rho_{\bm k}}{\partial
{\bm k}} + \frac{i}{\hbar}\;
[ {\cal{H}}^{(1)}_{\bm k}, \rho_{\bm k}] +
Q_{\bm k}\{ \rho \} = 0
\:,
\end{equation}
where $Q_{\bm k} \{ \rho \}$ is the collision integral and
${\cal{H}}^{(1)}_{\bm k}$ is the linear-${\bm k}$ Hamiltonian.
Similarly to the spin-galvanic effect there exist two different
mechanisms of the current-to-spin transformation, namely,
spin-flip mediated and precessional.

\subsection{Spin-Flip Mediated Current-Induced Polarization}
\begin{figure}[t]
\centering
\includegraphics*[width=7cm]{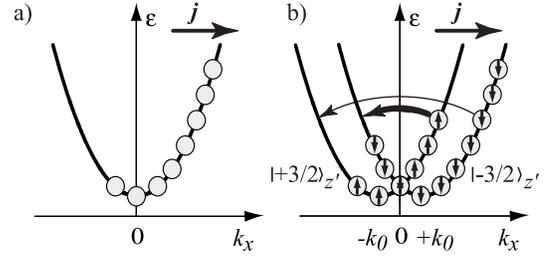}
\caption[]{Comparison of current flow in (a) spin-degenerate and (b)
spin-split subbands. (a)
%Electron
Hole distribution at a stationary
current flow due to acceleration in an electric field and momentum
relaxation. (b) Spin polarization due to spin-flip scattering.
Here only $\beta_{z'x} \sigma_{z'} k_x$ term is taken into account in the
Hamiltonian which splits the valence subband into two parabolas with
spin-up  $|+3/2\rangle_{z'}$  and spin-down $|-3/2\rangle_{z'}$ in the $z'$-direction. Biasing
along the $x$-direction causes an asymmetric in ${\bm k}$-space
occupation of both parabolas. 
{
%\color{blue}
After \cite{Ganichev04p0403641}}.
}
\label{figure12_currentspinmod}       % Give a unique label
\end{figure}

In the spin-flip mediated mechanism, a value of the
spin generation rate is calculated neglecting the commutator in
Eq.~(\ref{denkin}) and taking into account the
spin-flip processes in the collision integral and the linear-${\bm
k}$ terms in the electron dispersion. Microscopic illustration of
this mechanism is sketched in Fig.\,\ref{figure12_currentspinmod}(b) for
a 2D hole gas in a system of the C$_s$ symmetry, a situation
relevant for the experiments of Ref.~\cite{Ganichev04p0403641}. 

In the simplest case the electron kinetic energy in a QW depends
quadratically on the in-plane wave vector ${\bm k}$. In
equilibrium, the spin degenerate ${\bm k}$ states are
symmetrically occupied up to the Fermi energy $E_F$. If an
external electric field is applied, the charge carriers drift in
the direction of the resulting force. The carriers are accelerated
by the electric field and gain kinetic energy until they are
scattered, Fig.\,\ref{figure12_currentspinmod}(a). A stationary
state forms where the energy gain and the relaxation are balanced
resulting in an asymmetric distribution of carriers in the ${\bm
k}$-space. The
%electrons
holes acquire the average quasi-momentum
\begin{equation}\label{momentum}
\hbar \bar{\bm k}  = - e \tau_p {\bm F} = - {m_c
\over e n_s} {\bm j}\:,
\end{equation}
where $\tau_p$ is the momentum relaxation
time, ${\bm j}$ the
electric current density, $m_c$ the effective mass and $n_s$ the
2D carrier concentration. As long as the energy band is spin
degenerated in the ${\bm k}$-space a current is not accompanied by
spin orientation. However, in zinc-blende-lattice QWs or strained
bulk semiconductors the spin degeneracy is lifted due to the
linear-${\bm k}$ terms given by~Eq.~(\ref{lkterms}). To be
specific for the mechanism depicted in
Fig.\,\ref{figure12_currentspinmod}(b) we consider solely spin-orbit
interaction of the form $\beta_{z'x} \sigma_{z'} k_x$. Then the
parabolic energy band splits into two parabolic subbands of
opposite spin directions, $s_{z'} = 3/2$ and $s_{z'} = - 3/2$,
with minima symmetrically shifted in the ${\bm k}$-space along the
$k_x$ axis from the point $k = 0$ into the points $\pm k_0$, where
$k_0 = m_c \beta_{z'x}/ \hbar^2$. The corresponding dispersion is
sketched in Fig.\,\ref{figure12_currentspinmod}(b). 

In the presence of an in-plane electric field ${\bm F} \parallel x$ the
distribution of carriers in the ${\bm k}$-space gets shifted
yielding an electric current. Until the spin relaxation is
switched off the spin branches are equally populated and equally
contribute to the current. Due to the band splitting, spin-flip
relaxation processes $\pm 3/2 \to \mp 3/2$ are different because
of the difference in quasi-momentum transfer from initial to final
states. In Fig.\,\ref{figure12_currentspinmod}(b) the $\bm
k$-dependent spin-flip scattering processes are indicated by
arrows of different lengths and thicknesses. As a consequence
different amounts of spin-up and spin-down carriers contribute to
the spin-flip transitions causing a stationary spin orientation.
Thus, in this picture we assume that the origin of the current
induced spin orientation is, as sketched in
Fig.\,\ref{figure12_currentspinmod}(b), exclusively due to
scattering and hence dominated  by the Elliott-Yafet spin
relaxation processes.

% For figures use

\begin{figure*}[t]
\centering
\includegraphics*[width=11.5cm]{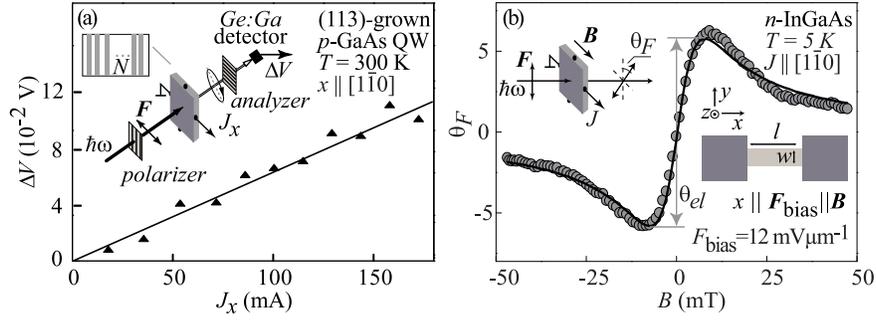}
\caption[]{ (a) Polarization dependent signal for
current in the active direction as a function of current strength for
two samples.  After Ref.~\protect\cite{Ganichev04p0403641}.
(b) Voltage-induced angle $\theta_F$ as a function
of the magnetic field $B$ for $F = 12$~mV ${\mu m}^{-1}$
(${\bm F} \parallel [1\bar{1}0])$.
After Ref. \protect\cite{Kato04p176601}. Open
circles are data, and lines are fits according Eq.~\protect(\ref{szsx}).
Insets to both panels show experimental set-up: (a) The sample is placed between
crossed polarizer and analyzer blocking optical transmission at
zero current through the sample. Injecting a modulated current in
the sample yields a signal at the detector which is recorded by the
box-car technique. (b) The current yields an in-plane spin polarization
which applying  magnetic field $B$ is rotated out off plane yielding Faraday rotation of the probe light. Second inset shows
sample geometry. Here dark areas are Ni/GeAu contacts and the light grey area is the InGaAs channel. }
\label{figure13_awshganichev}       
\end{figure*}

\subsection{Precessional Mechanism}
The precessional mechanism resulting in the current induced spin
orientation is based on the Dyakonov--Perel spin relaxation.
 In
this mechanism of spin polarization the contribution of spin-flip
scattering to the collision integral is ignored and the spin
appears taking into account the linear-${\bm k}$ Hamiltonian, both
in the collision integral and the commutator $[
{\cal{H}}^{(1)}_{\bm k}, \rho_{\bm k}]$. For example, we present
here the collision integral for elastic scattering
\begin{eqnarray} \label{qrho}
&&Q_{\bm k}\{ \rho\} = \frac{2 \pi}{\hbar} N_i \sum_{{\bm k}'}  \left|
A_{{\bm k}' {\bm k}} \right|^2 \\ &&\times \left\{ \delta \left(
E_{{\bm k}} + {\cal{H}}^{(1)}_{{\bm k}} - E_{{\bm k}'} -
{\cal{H}}^{(1)}_{{\bm k}'} \right) , \rho_{\bm k} - \rho_{{\bm
k}'} \right\}\:, \nonumber
\end{eqnarray}
where $E_{\bm k} = \hbar^2 k^2/(2 m_c)$, $N_i$ is the density of
static defects acting as the scatterers, $A_{{\bm k}' {\bm k}}$ is
the scattering matrix element and the braces mean the
anticommutator, $\{ A B \} = (AB + BA)/2$ for two arbitrary
2$\times$2 matrices $A$ and $B$. Similar equation can be written
for electron-phonon scattering.

In the equilibrium the electron spin density matrix is given by
\begin{equation}
\rho^0_{\bm k} = f^0 (E_{\bm k} + {\cal{H}}^{(1)}_{{\bm k}}) \approx
f^0 (E_{\bm k}) + \frac{\partial f^0}{\partial E_{\bm k}} {\cal{H}}^{(1)}_{{\bm k}} \:,
\end{equation}
where $f^0(E) = \{ \exp{[(E - \mu)/k_B T]} + 1 \}^{-1}$ is the
Fermi-Dirac distribution function, $\mu$ is the electron chemical
potential, $k_B$ is the Boltzmann constant and $T$ is the
temperature.

Neglecting the spin splitting we can write the solution of
Eq.~(\ref{denkin}) in the text-book form
\begin{equation} \label{shiftF}
f_{\bm k} = f^0 (E_{\bm k}) - eF_x v_x \tau_1(E_{\bm k})
\frac{\partial f^0}{\partial E_{\bm k}}
\end{equation}
with ${\bm s}_{\bm k} = 0$. Here $v_x = \hbar k_x/m_c$, $\tau_1$
is the time describing the relaxation of a distribution-function
harmonic with the angular dependence of the functions $k_x$ or
$k_y$. If we substitute $\rho_{\bm k}$ with the distribution
function (\ref{shiftF}) into the collision integral
and $\rho^0_{\bm k}$ into the first term of Eq.~(\ref{denkin}) we
obtain an equation for ${\bm s}_{\bm k}$. By solving this equation
one arrives at the estimation for the spin density
\begin{equation} \label{finaleq}
s_{\mu} \equiv \sum\limits_{\bm k} {\bm s}_{\bm k, \mu} \sim
\beta_{\mu \lambda} \bar{k}_{\lambda} g_{2d} \:,
\end{equation}
where $g_{2d} = m_c/(\pi \hbar^2)$ is the 2D density of states and
$\hbar \bar{k}_{\lambda} / m_c$ is the electron drift velocity.
The factor of proportionality can be found in \cite{VaskoRai,Raichev,GolIvch,Raimondi2016,Gorini2017}.

Spin orientation by electric current in low-dimensional structures has been observed applying
various experimental techniques, comprising transmission of
polarized THz-radiation, polarized luminescence and space resolved
Faraday 
rotation~\cite{Yang06,Ganichev04p0403641,Silov2004,Kato04p176601,Silov04,
Sih05isge,Stern06isge,Beschoten,Awschalom}.
Here we briefly sketch results of experiments on the
THz-transmission and polarized photoluminescence in which the spin
orientation by electric current in QW structures was initially
observed.

It has been shown that in the streaming regime the current-induced spin orientation is remarkably increased \cite{streaming}. A microscopic theory of spin orientation by electric current  for the hopping regime has been developed by Smirnov and Golub \cite{hopping}.

\subsection{Current Induced Spin Faraday Rotation
}
In order to observe current induced spin polarization,
in~\cite{Ganichev04p0403641} the circular dichroism and Faraday
rotation  of THz radiation transmitted through samples containing
multiple QWs were studied.
This method allows one to detect spin polarization
at normal incidence in the growth direction. The materials chosen
for studies were (113)- and miscut (00l)-oriented $p$-type GaAs
multiple QWs of the C$_s$ point group symmetry.
The transmission measurements were carried out at room temperature
using  linearly polarized $\lambda=118~\mu$m radiation as shown in
Fig.~\ref{figure13_awshganichev}(a): the sample is placed between
two metallic grid polarizers and the $cw$-terahertz radiation is
passed through this optical arrangement. 

Using modulation technique the Faraday rotation was observed only for
the current flowing in the
$x$-direction. This is in agreement with the phenomenological
equation $S_{z'} = R_{z'x} j_x$ relating the induced spin with the
current density, the spin polarization can be obtained only for
the current flowing along the direction normal to the mirror
reflection plane which is perpendicular to the
$x$-axis. The signal $\Delta V$ caused by rotation of polarization
plane is shown in Fig.~\ref{figure13_awshganichev}(a) as a function
of the current strength. Experiment shows that, in agreement  with
Eq.~(\ref{finaleq}), the spin polarization increases
with the decreasing temperature.

Current induced spin polarization has also been detected by the
Faraday rotation of infrared radiation applying a mode-locked
Ti:sapphire laser. Figure~\ref{figure13_awshganichev}(b)
demonstrates an optical detection of current-induced electron spin
polarization in strained InGaAs epitaxial
layers~\cite{Kato04p176601}. The heterostructure studied consists
of 500 nm of $n$-In$_{0.07}$Ga$_{0.93}$As (Si doped for $n = 3
\times 10^{16}$ cm$^{-3}$) grown on (001) semiinsulating GaAs
substrate and capped with 100 nm of undoped GaAs. The $n$-InGaAs
layer is strained due to the lattice mismatch. An alternating
electric field ${\bm F}$ is applied along either of the two
crystal directions [110] and $[1\bar{1}0]$, the in-plane magnetic
field ${\bm B}$ is parallel to ${\bm F}$. A linearly polarized
probe beam is directed along the $z$ axis, normally incident and
focused on the sample. The polarization axis of the transmitted
beam rotates by an angle $\theta_F$ that is proportional to the
$z$ component of the spins $S_z$ (the spin Faraday rotation). 

The current-induced angle is lock-in detected at the modulation
frequency as a function of the applied magnetic field.
The experiment data in Fig.~\ref{figure13_awshganichev}(b) can be
explained by assuming a constant orientation rate $\dot{s}_{{\bm
k}, y}$ for spins polarized along the $y$ axis. The rotation of
the spins around the magnetic field yields the $z$ spin component
given by
\begin{equation} \label{szsx}
S_z (B) =  \frac{ \omega_L \tau_{s \parallel} }{1 + (\omega_L \tau_s )^2}\:
S_{0y}\:,\: S_{0y} = \tau_{s, \perp} \sum_{\bm k} \frac{\dot{s}_{{\bm k}, y}}{n_s}\:,
\end{equation}
where the notations for the spin relaxation times and Larmor
frequency are introduced in
Eq.~(\ref{Hanle}). The high sensitivity of the Faraday rotation
technique allows detection of 100~spins in an integration time of
about 1~s, unambiguously revealing the presence of a small spin
polarization due to laterally applied electric fields.

% For figures use
\begin{figure}[t]
\centering
\includegraphics*[width=5.7cm]{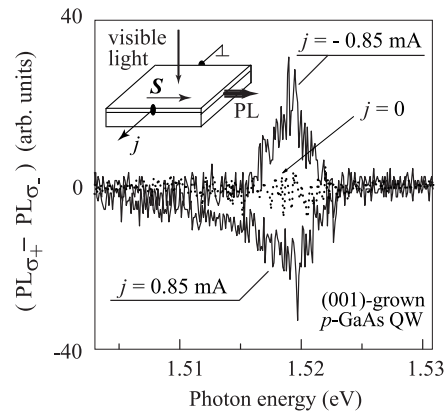}
\caption[]{Differential spectra of polarized PL for two current directions.
{
%\color{blue}
After\protect\cite{Silov2004}}. Base line  is
taken with the current turned off. Inset shows experimental geometry.}
\label{figure14_silov}       % Give a unique label
\end{figure}

\subsection{Current Induced Polarization of Photoluminescence}
In~\cite{Silov2004,Silov04} in order to detect the
inverse spin-galvanic effect
the degree of circular polarization of the 2D hole gas
photoluminescence (PL) was measured.
This experimental procedure has become a proven method for probing
spin polarization\cite{Meier,spintronicbook02}. A (001)-grown
sample cleaved into bars was studied with the current flowing
along the long side cleaved parallel to the
[1$\bar{1}$0] direction. Lately (113)-grown samples were also
studied~\cite{Silov04}. The PL was excited with 633\,nm line from
a helium-neon laser. In (001)-oriented samples the PL was
collected from the cleaved (110) facet of the sample. 

On the other hand, at the (113)-oriented heterojunctions, because of the C$_s$
symmetry, the mean spin density will have a component along the
growth direction. Therefore, the PL in this case was detected in
the back scattering geometry; the degree of PL circular
polarization $P_{\rm c}$ was analyzed with a $\lambda/4$ plate and
a linear polarizer. Inset in Fig.\,\ref{figure14_silov} shows the
experimental arrangement for measuring the current induced
polarization and differential spectra, (PL$_{\sigma_+}$ -
PL$_{\sigma_-}$), for the two opposite current directions. The
observation of the circularly polarized radiation and, in
particular, the reversal of helicity upon the inversion of the
current direction demonstrate the effect of current induced spin
polarization. The observed degree of polarization in (001)-grown
samples yields a maximum of 2.5\%~\cite{Silov2004}. In (113)-grown
samples even higher polarization of 12\% at 5.1 K is
achieved~\cite{Silov04}.

\section{Pure Spin Currents}
Pure spin current represents a
nonequilibrium distribution where free carriers, electrons or
holes, with the spin up propagate mainly in one direction and
an equal number of spin-down carriers propagates in
the opposite direction. This state is characterized by zero charge
current because electric currents contributed by spin-up and
spin-down quasi-particles cancel each other, but leads to
separation of spin-up and spin-down
electron spatial distributions and accumulation of the opposite
spins at the opposite edges of the sample. Spin currents in
semiconductors can be driven by an electric field acting on
unpolarized free carriers (the spin Hall effect, not considered
here). They can be induced as well by
optical means under interband or intraband optical transitions in
noncentrosymmetric bulk and low-dimensional
semiconductors~\cite{Bhat,Tarasenko05p292,Zhao,purespin2,Ganichev06zerobias,PRBSiGe}.

\subsection{Pure Spin Current Injected by a Linearly Polarized Beam}
In general, the spin current density pseudotensor
$q_{\lambda \mu}$ describes the flow of the $\mu$
component of the spin polarization in the spatial direction
$\lambda$. Phenomenologically, the spin photocurrent
$q_{\lambda \mu}$ is related with bilinear products
$I e_{\nu} e^*_{\eta}$ by a fourth-rank tensor, in
Eq.~(\ref{psc}) it is the tensor $P_{\lambda \mu \nu \eta}$. Here
we assume the light to be linearly polarized. In this particular
case the product $e_{\nu} e^*_{\eta} \equiv e_{\nu} e_{\eta}$ is
real and the tensor $P_{\lambda \mu \nu \eta}$ is symmetric with
respect to interchange of the third and fourth indices.

Among microscopic mechanisms of the pure spin photocurrent we
first discuss those related to the ${\bm k}$-linear terms in the
electron effective Hamiltonian~\cite{Tarasenko05p292,purespin2}. Let us
consider the $e1$-$hh1$ interband absorption of linearly polarized
light under normal incidence on (001)-grown QWs. In this case
linear-${\bm k}$ splitting of the $hh1$ heavy-hole valence subband
is negligibly small. For the sake of simplicity, but not at the
expense of generality, in the $e1$ conduction subband we take into
account the spin-orbit term $\beta_{yx} \sigma_y k_x$ only, the
contribution to the tensor ${\bm P}$ coming from the term
$\beta_{xy} \sigma_x k_y$ is considered similarly. Then the
conduction-electron spin states are eigen states of the spin
matrix $\sigma_y$. For the light linearly polarized, say, along
$x$, all the four transitions from each heavy-hole state $\pm 3/2$
to each $s_y = \pm 1/2$ state are allowed. The energy and momentum
conservation laws read
\[
E_g^{\rm QW} + \frac{\hbar^2 (k_x^2 + k_y^2)}{2 \mu_{cv}} + 2
s_y \beta_{yx} k_x = \hbar \omega\:,
\]
where we use the same notations for $E_g^{\rm QW}$ and $\mu_{cv}$
as in Subsection~\ref{interbandoptical}. For a fixed value of
$k_y$ the photoelectrons are generated only at two values of $k_x$
labeled $k^{\pm}_x$. The average electron velocity in the $s_y$
spin subband is given by
$$
\bar{v}_{e,x}= \frac{\hbar(k^{+}_x
+ k^{-}_x)}{2m_c} + 2 s_y \frac{\beta_{yx}}{\hbar}=
\frac{2 s_y \beta_{yx}}{\hbar} \frac{m_c}{m_c + m_v} \:.
$$

The   resulting  spin fluxes ${\bm i}_{\pm 1/2}$ are opposite in sign, the
electric current ${\bm j} = e ({\bm i}_{1/2} + {\bm i}_{- 1/2})$
is absent but the spin current 
{
%\color{blue}
${\bm j}_s = ({\bm i}_{1/2} -
{\bm i}_{- 1/2})/2$} is nonzero. This directional movement decays in
each spin subband within the momentum relaxation time
$\tau^e_p$. However under the cw
photoexcitation the electron generation is continuous which
results in the spin current
\[
q_{xy} = \frac{\beta_{yx} \tau^e_p}{2 \hbar} \frac{m_c}{m_c + m_v}
\frac{\eta_{eh} I}{\hbar \omega}\:.
\]
Under the normal incidence it is independent of the light
polarization plane.

 Now we turn to the pure spin currents 
excited in (110)-grown QWs. In these structures the spin component along the normal
$z'\|[110]$ is coupled with the in-plane electron wave vector ${\bm k} \parallel [1\bar{1}0]$ due
to the term $\beta_{z' x} \sigma_{z'}k_x$ in the conduction band
and the term proportional to $J_{z'} k_x$ in the heavy-hole band
where $J_{z'}$ is the 4$\times$4 matrix of the angular momentum
3/2. The coefficient $\beta_{z' x}^{(e1)}$ is relativistic and can
be ignored compared with the nonrelativistic constant
$\beta_{z'x}^{({\rm hh1})}$ describing the spin splitting of
heavy-hole states. The allowed direct optical transitions from the
valence subband $hh1$ to the conduction subband $e1$ are $|+3/2
\rangle \rightarrow |+1/2 \rangle$ and $|-3/2 \rangle \rightarrow
|-1/2 \rangle$, where $\pm 1/2, \pm 3/2$ indicate the $z'$
components of the electron spin and hole angular momentum. Under
linearly polarized photoexcitation the charge photocurrent is not
induced, and for the electron pure spin photocurrent one has
\begin{equation} \label{fz'x}
q_{x z'} = \frac{\beta_{z'x}^{(\rm hh1)} \tau^e_p}{2 \hbar} \frac{m_v}{m_c + m_v}
\frac{\eta_{eh} I}{\hbar \omega}\:.
\end{equation}
The similar hole spin current can be ignored in the spin
separation experiments because of the much
shorter spin relaxation time for holes as compared to the
conduction electrons.

Another contribution to spin photocurrents comes from ${\bm
k}$-linear terms in the matrix elements of the interband optical
transitions~\protect\cite{TarICPS}. Taking into account ${\bm
k}\cdot{\bm p}$ admixture of the remote $\Gamma_{15}$ conduction
band to the $\Gamma_{15}$ valence-band states $X_0({\bm k}),
Y_0({\bm k}), Z_0({\bm k})$ and the $\Gamma_{1}$ conduction-band
states $S({\bm k})$, one derives the interband matrix elements of
the momentum operator for bulk zinc-blende-lattice
semiconductors~\cite{voisin,Khurgin}
\begin{equation}
\langle i S ({\bm k}) | {\bm e} \cdot{\bm p}| X_0({\bm k}) \rangle =
{\cal P} [e_{x_0} + i \chi (e_{y_0} k_{z_0} + e_{z_0} k_{y_0})] \:,\\
\end{equation}
$\langle i S({\bm k}) | {\bm e} \cdot{\bm p}| Y_0 ({\bm k})
\rangle $ and $\langle i S({\bm k}) | {\bm e} \cdot{\bm p}|
Z_0({\bm k}) \rangle$ are obtained by the cyclic permutation of
indices, the coefficient $\chi$ is a material parameter
dependent on the interband spacings and interband
matrix elements of the momentum operator at the $\Gamma$ point,
and we use here the crystallographic axes $x_0 \parallel [100]$
etc. For GaAs band parameters~\cite{Jancu} the coefficient $\chi$
can be estimated as $0.2$~\AA. 

 Calculation for (110)-grown QWs shows that the spin photocurrent caused by ${\bm k}$-linear terms in the interband matrix elements has the form
\begin{eqnarray} \label{fz'xm}
&&q_{x z'} = \varepsilon (e_{y'}^2-e_{x}^2)
\frac{\chi \tau^e_p}{\hbar} \frac{\eta_{cv}}{\hbar\omega} I \:, \\
&&q_{y'z'} = \varepsilon e_{x} e_{y'} \frac{\chi \tau^e_p}{\hbar}
\frac{\eta_{ev}}{\hbar\omega} I \nonumber
\end{eqnarray}
with $\varepsilon=(\hbar\omega-E^{QW}_g)m_v/(m_c+m_v)$ being the
kinetic energy of the photoexcited electrons and $y' \parallel [0
0 \bar{1}]$. In contrast to Eq.~(\ref{fz'x}), this contribution
depends on the polarization plane of the incident light and
vanishes for unpolarized light. From comparison of
Eqs.~(\ref{fz'x}) and (\ref{fz'xm}) one can see that depending on
the value of $\hbar \omega - E^{QW}_g$ the two contributions to
$q_{xz'}$ can be comparable or one of them can dominate over the
other.

\begin{figure}[t]
\centering
\includegraphics*[width=7.5cm]{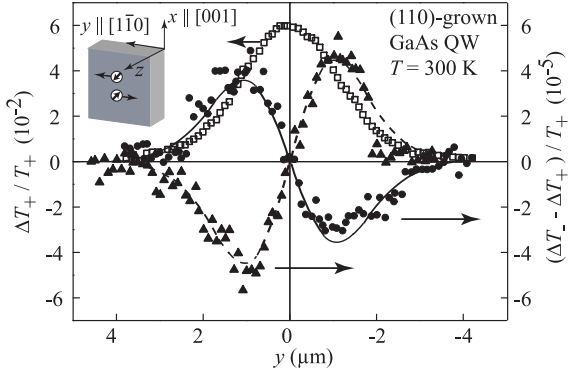}
\caption[]
{  Measurement of $\Delta T_+ /T_+ \propto n_{- 1/2}(y)$
(open squares) and $\Delta T_- /T_+ - \Delta T_+ /T_+ \propto
n_{1/2}(y) - n_{- 1/2}(y)$ for an $x$ polarized (solid circles)
and a $y$ polarized (solid triangles) pump pulse at room
temperature for a pump fluence of 10 $\mu$J/cm$^2$. The lines are the
fits to the data for spin separation
 $d = 2.8$~nm. After
\cite{Zhao}.}
\label{figure15_smirl_pure}
\end{figure}

The injection and control of pure spin currents in (110)-oriented GaAs QWs at room temperature by
one-photon absorption of a linearly polarized optical pulse was
demonstrated by Zhao et al.~\cite{Zhao}. Spatially resolved
pump-probe technique was used. The pump pulse excited electrons
from the valence to the conduction band with an excess energy of
$\sim 148$~meV large enough for the polarization-dependent
contribution (\ref{fz'xm}) to dominate over the
polarization independent contribution (\ref{fz'x}).
The probe was tuned near the band edge. The $\sigma_+$ component
of the linearly polarized probe interacts stronger with the
spin-down electrons of the density $n_{-1/2}$, while the
$\sigma_-$ component interacts stronger with the spin-up electrons
of the density $n_{1/2}$. Consequently, the net spin polarization
of the carriers present in the sample at the position of the probe
can be readily deduced from the difference in the transmission
$T_{\pm}$ of the $\sigma_+$ and $\sigma_-$ components of the
probe. 

The results of measuring $n_{- 1/2}(y) \propto \Delta T_+
/T_+$ and $\Delta n(y) \equiv n_{1/2}(y) - n_{- 1/2}(y) \propto
\Delta T_- /T_+ - \Delta T_+ /T_+$ for the $x$ and $y$ polarized
pump are shown in Fig.~\ref{figure15_smirl_pure}. Note that here
we retain the notations $x \parallel [001], y
\parallel [1 \bar{1} 0], z
\parallel [110]$ as they are introduced in the original paper \cite{Zhao}
while in Eqs.~(\ref{fz'x}) and (\ref{fz'xm}) we use the Cartesian
frame $x
\parallel [1 \bar{1} 0], y' \parallel [0 0 \bar{1}], z' \parallel
[110]$. Clearly, the $\Delta
n(y)$ signal is consistent with a pure spin current. It can be
well fitted by the product of the spatial derivative of the
original Gaussian profile and the separation $d$ of the order of
the photoelectron mean free path. The solid curve in
Fig.~\ref{figure15_smirl_pure} corresponds to a fit for $d$ = 28
\AA. In agreement with Eq.~(\ref{fz'xm}) the $\Delta n(y)$ signal
has opposite signs for the $x$ and $y$ linear polarization of the
pump.

It is worth to add that a pure spin current may be generated at
simultaneous one- and two-photon coherent excitation of proper
polarization as demonstrated in bulk GaAs~\cite{Stevens02p4382}
and GaAs/AlGaAs QWs~\cite{Stevens2003}. This phenomenon may be
attributed to a photogalvanic effect where the reduced symmetry is
caused by the coherent two-frequency
excitation~\cite{Entin89p664}.

\subsection{Pure Spin Currents Due to Spin-Dependent Scattering}
Light absorption by free carriers, or the Drude-like absorption,
is accompanied by electron scattering by acoustic or optical
phonons, static defects etc. Scattering-assisted photoexcitation
with unpolarized light also gives rise to a pure spin
current~\cite{Ganichev06zerobias,PRBSiGe}. However, in contrast to
the direct transitions considered above, the spin splitting of the
energy spectrum leads to no essential contribution to the spin
current induced by free-carrier absorption. The more important
contribution comes from asymmetry of the electron spin-conserving
scattering. In gyrotropic low-dimensional structures, spin-orbit
interaction adds an asymmetric spin-dependent term to the
scattering probability. This term in the scattering matrix element
is proportional to components of $[ {\bm \sigma} \times ({\bm k} +
{\bm k}')]$, where ${\bm \sigma}$ is the vector composed of the
Pauli matrices, ${\bm k}$ and ${\bm k}'$ are the initial and
scattered electron wave vectors.

\begin{figure*}[t]
\centering
\includegraphics*[width=10.5cm]{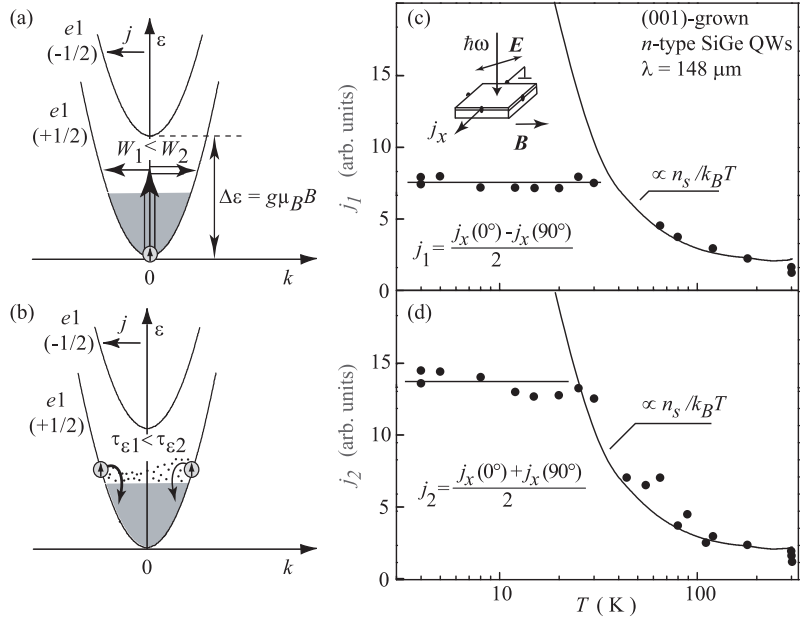}
\caption[]
{ Microscopic origin of a zero-bias spin separation and the
corresponding magnetic field-induced photocurrent for excitation (a) and relaxation (b) models
corresponding to currents $j_1$ and $j_2$, see explanation
in the text.  Temperature dependencies of the
contributions $j_1$ (c) and $j_2$ (d) to the photocurrent $j_x$ in a
magnetic field ${\bm B} \parallel y$.
Full lines are fits to $A n_s /k_BT$
with a single fitting parameter $A$,
and to a constant, respectively.
{
%\color{blue} 
After}  \cite{PRBSiGe}.
}
\label{figure16_MGEcombined}
\end{figure*}

Figure~\ref{figure16_MGEcombined}(b) sketches the process of
energy relaxation of hot electrons for the spin-up subband
($s=+1/2$)  in a quantum well containing a 2D
electron gas. Energy relaxation processes are shown by
curved arrows.
Due to the spin-dependent scattering, transitions to positive and
negative $k'_x$ states occur with different probabilities. In
Fig.~\ref{figure16_MGEcombined}(b) the difference is
indicated by curved arrows of different thickness.
This asymmetry causes an imbalance in the distribution of carriers
in both subbands ($s=\pm 1/2$) between positive and negative
$k_x$-states.
This in turn yields a net electron flows, $\bm{i}_{\pm 1/2}$,
within each spin subband. Since the asymmetric part of the
scattering amplitude depends on spin orientation, the
probabilities for scattering to positive or negative
$k^\prime_x$-states are inverted for spin-down and spin-up
subbands.
Thus, the charge currents, $\bm{j}_+ = e\bm{i}_{+1/2}$ and
$\bm{j}_- = e\bm{i}_{-1/2}$, where $e$ is the electron charge,
have opposite directions because $\bm{i}_{+1/2} = -\bm{i}_{-1/2}$
and therefore they cancel each other. Nevertheless, a finite pure
spin current 
{
%\color{blue}
$\bm{j}_s = (\bm{i}_{+1/2} -
\bm{i}_{-1/2})/2$} is generated since electrons with spin-up and
spin-down move in opposite directions
\cite{Ganichev06zerobias}.

Similarly to the relaxation mechanism, optical
excitation of free carriers by Drude absorption, also involving
electron scattering, is asymmetric and yields spin separation.
Figure~\ref{figure16_MGEcombined}(a) sketches the process of Drude
absorption via virtual states for the spin-up subband. Vertical
arrow indicates optical transitions from the initial state with
$k_x = 0$ while the horizontal arrows describe an elastic
scattering event to a final state with either positive or negative
electron wave vector. Due to the spin dependence of scattering,
transitions to positive and negative $k_x$ states occur with
different probabilities. This is indicated by the different
thickness of the horizontal arrows. The asymmetry causes an
imbalance in the distribution of photoexcited carriers in the spin
subband between positive and negative $k_x$-states. This in turn
yields electron flow.

\subsubsection{Magneto-Gyrotropic Effects
}
\label{MGEf}
A pure spin current and zero-bias spin separation can be converted
into a measurable electric current by application of a magnetic
field. Indeed, in a Zeeman spin-polarized system, the two fluxes
${\bm i}_{\pm 1/2}$, whose magnitudes depend on the free carrier
densities in spin-up and spin-down subbands, $n_{\pm 1/2}$,
respectively, do no longer compensate each other and hence yield a
net electric current. Since the fluxes ${\bm i}_{\pm 1/2}$ are
proportional to the carrier densities $n_{\pm 1/2}$ the charge
current is given by
\begin{equation}
{\bm j} = e ({\bm i}_{1/2} + {\bm i}_{-1/2}) = 4 e S {\bm j}_s\:,
\end{equation}
where $x \parallel [1\bar{1}0], y \parallel [110]$,
$S = (1/2) (n_{1/2} - n_{- 1/2})/ (n_{1/2} + n_{- 1/2})$ is
the average spin per particle and ${\bm j}_s$ is the
pure spin current in the absence
of magnetic field. An external magnetic field ${\bm
B}$ results in different equilibrium populations of the two spin
subbands due to the Zeeman effect. We remind
that in equilibrium the average spin is given by 
%\color{blue}
${\bm S} = - g
\mu_B {\bm B}/ 4 k_B T $ for a nondegenerate 2D electron gas and ${\bm S} = - g
\mu_B {\bm B}/ 4 E_F $ for a degenerate one, where $\mu_B$ is the Bohr magneton and $g$ is the electron $g$-factor (or Land\'e factor).

In the structures of the C$_{2v}$ symmetry, the phenomenological
equation~(\ref{mge}) for the magneto-photogalvanic
effects
induced by normally-incident linearly polarized radiation reduces
to~\cite{Belkov05p3405}
\begin{eqnarray}
\label{phen} j_{x} = S_1  B_{y}I + S_2 B_{y}
\left( e_{x}^2 - e_{y}^2 \right) I + 2 S_3
B_{x} e_{x} e_{y} I \:,\\
j_{y} = S'_1  B_{x}I + S'_2 B_{x}  \left(
e_{x}^2 - e_{y}^2 \right) I+ 2S'_3 B_{y}
e_{x} e_{y} I  \:. \nonumber
\end{eqnarray}
Here the parameters $S_1$ to $S_3$ and $S^\prime_1$ to
$S^\prime_3$ are linearly independent components of the tensor
$\Phi_{\lambda \mu \nu \eta}$ in Eq.~(\ref{mge}) and only in-plane
components of the magnetic field are taken into account. For ${\bm
B} \parallel y$ we have
\begin{equation} \label{NatPhys}
j_{x} = j_1 \cos{2 \alpha} + j_2 \:,\: j_y = j_3 \sin{2 \alpha}\:,
\end{equation}
where $j_1 = S_2 B_{y} I$, $j_2 = S_1 B_{y} I$, $j_3 = S'_3 B_{y}
I$.

Right panels of Fig.~\ref{figure16_MGEcombined}  show the
temperature dependence of the currents $j_1$ and $j_2$
corresponding to the excitation and relaxation mechanisms depicted
in Figs.~\ref{figure16_MGEcombined}(c) and
\ref{figure16_MGEcombined}(d), respectively. The data are obtained
in an $n$-type SiGe QW structure for the magnetic field 0.6 T
under excitation with a THz molecular laser ($\lambda = 148$
$\mu$m). The analysis shows~\cite{Ganichev06zerobias,PRBSiGe} that the temperature
dependence of the current can be reduced to $n_s S$, the current
becomes independent of temperature at low temperatures and is
proportional to $n_s/T$ at high temperatures, in agreement with
the experimental data. Thus, the application of an external
magnetic field gives experimental access to investigations of pure
spin currents. 
{
%\color{blue} 
The conversion of the pure spin current has been demonstrated for GaAs, InAs, InSb and Cd(Mn)Te QWs, for review see~\cite{Belkovbook}, as well as for Dirac fermions in HgTe QWs of critical thickness~\cite{Olbrich2012}.} Like circular PGE and spin-galvanic effect, magneto-gyrotropic effect provides an efficient tool for investigation of inversion asymmetry as it is demonstrated for (110)-grown GaAs quantum wells \cite{67}. 

\subsection{Spin vs. Orbital Mechanisms of Photocurrents}

{
%\color{blue} 
While analyzing measured data on the polarization-dependent magneto-photogalvanic effects an attention should be payed to possible orbital mechanisms of the photocurrent formation \cite{spivakx,Torbital,Orbital2010,Taras2011,TarasNature,nanowireSi,Falko}. Spin and orbital mechanisms may simultaneously be responsible  for various physical phenomena and result in two competitive contributions to the observables. The textbook example is the Pauli paramagnetism and the Landau diamagnetism that yield comparable contributions to the magnetic susceptibility of an electron gas. 

In addition to the spin mechanisms reviewed in this article, the photocurrents induced by circularly or linearly polarized radiation may also steam from orbital effects caused by quantum interference of optical transitions~\cite{circSi}, ratchet effects in QWs with a lateral superlattice~\cite{orbital0}, dynamic Hall effect~\cite{orbital1}, pure valley currents~\cite{orbital2}, magnetic field dependent scattering~\cite{orbital3} etc. It is worth recalling that, in the 
first papers~\cite{Belinicher,Ivchenko78p640} on the circular PGE, Belinicher proposed an orbital mechanism of the effect whereas Ivchenko and Pikus predicted the  spin-dependent circular photocurrent related to linear ${\bm k}$-terms in the free-carrier effective Hamiltonian. In materials with vanishingly small spin-orbit interaction, like Si or graphene, these mechanisms are obviously predominant.

An interplay of spin and orbital mechanisms has been demonstrated in GaAs/Al$_x$Ga$_{1-x}$As
QW structures~\cite{gFactorSign} where the former contribution is proportional to the nonequilibrium spin while the latter is caused by a magnetic-field-induced and spin-independent scattering asymmetry. To separate them, the well-known fact that in GaAs based
QWs the Zeeman splitting changes its sign at a certain QW
width~\cite{Ivchenkobook2} has been utilized. This inversion is mostly caused by the opposite
signs of the $g$-factor 
in bulk GaAs and AlGaAs and the deeper penetration of the electron wave function into the barrier with narrowing the QW width.

The experiment shows that for the most
QW widths the magneto-gyrotropic PGE is mainly driven by spin-related mechanisms
which result in a photocurrent proportional to the $g$-factor. For structures
with a vanishingly small $g$-factor, i.e. for QW thicknesses close to the $g$-factor inversion point, the PGE caused by orbital mechanisms is clearly observed. 

Experiments on (Cd,Mn)Te/(Cd,Mg)Te diluted magnetic semiconductor QWs~\cite{diluted} provide
even more spectacular evidence for the zero-bias spin separation
 as a cause of
the magneto-gyrotropic 
PGE. Diluted magnetic semiconductors
(DMS)
 are traditionally defined as diamagnetic semiconductors doped
with a few to several atomic percent of some transition metal with unpaired
$d$ electrons. Exchange interaction between the localized electrons of $d$ shells of
the magnetic ions and delocalized band carrier states gives rise to specific
features of DMS, e.g., the exchange enhanced Zeeman splitting of the electronic bands
and the giant Faraday rotation. Naturally, the conversion of the zero-bias spin separation into the net electric current is also strongly affected by the magnetic properties of DMS structures. 

The spin polarization of magnetic ions in an external magnetic field ${\bm B}$ not only enhances the spin photocurrent due to the exchange enhanced Zeeman effect but also disturbs the balance between electron 
flows with opposite spins due to spin-dependent scattering by localized magnetic ions. 
The Zeeman splitting is given by a sum of intrinsic and exchange contributions of opposite signs, the former being independent of temperature and the latter being proportional to the  modified Brillouin function $B_{5/2}(\xi)$, where $$\xi =  \frac52 \frac{g_{\rm Mn} \mu_B B}{k_B(T_{\rm Mn} + T_0)}\:,$$
$g_{\rm Mn}$$=$$\mbox{}\;2$, $T_{\rm Mn}$ is the Mn-spin subsystem temperature and $T_0$ accounts for the Mn-Mn antiferromagnetic interaction.
The strong temperature dependence of the exchange contribution governs the temperature
behaviour of the spin photocurrent. In particular, it results in a reversal of
the photocurrent direction with temperature variation. Experiments on Mn-doped II-VI, III-V, and hybrid III-V/II-VI QW structures excited by microwave or terahertz
radiation have demonstrated that, as well as in other spin-dependent phenomena in DMS, the strength of photocurrent can be widely tuned by temperature,
magnetic field and concentration of the magnetic ions.}

\section{Photocurrents of Dirac Fermions in Topological Insulators}
In recent years, much attention in condensed-matter physics is directed towards studies of electronic properties of Dirac fermions in three-dimensional (3D) and 2D topological 
insulators 
(TI) which
are nonmagnetic insulators in the ``bulk'' but have robust gapless edge/surface states described by a Dirac equation for massless particles. 

Many materials have been proposed already as TIs, including
the 3D  family Bi$_2$Se$_3$, Bi$_2$Te$_3$, Sb$_2$Te$_3$ and their alloys, an alloy of Bi and Sb, narrow-gap semiconductor  HgTe/CdTe and AlSb/InAs/GaSb/AlSb quantum wells, strained HgTe films as well as topological crystalline insulators Pb$_{1-x}$Sn$_{x}$Se~\cite{Hasan, Ando}. 
They are a new playground for many interesting phenomena in the spin physics. In particular, it has been demonstrated that the surface Dirac electron states have helical spin structure and possess a strong relationship between the charge current and spin polarization. The spin-dependent photogalvanic effects, direct and inverse, could not be better suited for establishing this 
relation~\cite{inverseTI,HosurBerry,TopIns4,Ultrathin,TopIns2,TopIns5,TItheory,TopIns3dan,Ultrafast,TopIns1,TopIns2add,TopIns3,TopIns2add2}.

\begin{figure}[t]
\centering
\includegraphics*[width=8.5cm]{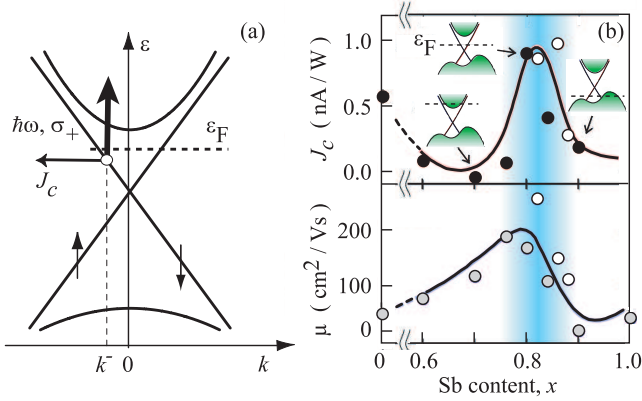}
\caption[]
{ {
%\color{blue}
({\bf a}) Schematic picture of optical transitions from helical edge states (straight lines marked by upward and downward arrows show spin-up and spin down branches, respectively)
to bulk conduction band states (upper parabola). The vertical arrow illustrates asymmetry of the photoexcitation by circularly polarized radiation $\sigma_+$: the probability of edge-bulk transition is spin-dependent and different for the spin-up and spin-down branches.
Horizontal arrow sketches the resulting circular photocurrent, $J_c$.
({\bf b}) The circular PGE current as a function of the Sb content $x$ plotted together with the mobility. The (Bi$_{1-x}$Sb$_x$ )$_2$Te$_3$ samples with the data represented with open circles were fabricated in a different chamber condition. The solid lines are guides to the eyes. The shaded area indicates the compositions where the Fermi energy $\varepsilon_{\rm F}$ is in the bulk band gap. The insets show the position of $\varepsilon_{\rm F}$ for three particular samples.}
{
%\color{blue} 
After \cite{TopIns1}.}}
\label{fig_ti1}
\end{figure}

McIver et al.~\cite{TopIns4} have shown that the illumination of the topological insulator Bi$_2$Se$_3$ with circularly polarized light generates a photocurrent originating from topological helical Dirac fermions and that the reversal of the light helicity reverses the direction of the photocurrent.
Later this effect has been observed for helical edge states of HgTe-based 2D TIs~\cite{TopIns2add2}. Furthermore, Olbrich et al. demonstrated that, by contrast to conventional QW structures discussed above, the spin-dependent photogalvanic effects in TI systems can even be generated by absorption of linearly polarized radiation \cite{TopIns2}.

An important advantage of the spin photogalvanics is that it, being forbidden by symmetry in most 3D TIs, can be used to probe selectively  surface 
states
 of novel materials. This is particularly helpful in the search for room temperature 3D TI where transport experiments are often handicapped by a large residual bulk charge carrier density.
{
%\color{blue}

Photogalvanic currents originating from 2D surface states or one-dimensional helical edge states can be excited by different kinds of optical transitions
including those between the edge and bulk states, the Drude absorption or the direct optical transitions between spin-up and spin-down branches of edge states characterized by linear dispersion.
Figure \ref{fig_ti1}(a) depicts the dispersion of helical dispersion of the edge states and the bulk conduction- and valence-band states 
(up and down parabolas).  The edge/bulk mechanism of circular PGE~\cite{TopIns4,TItheory} schematically shown in Fig.~\ref{fig_ti1}(a) implements similar physical concepts as used for the photocurrents arising due 
to inter-subband optical transitions and discussed in Sect.~\ref{ch7cpge}. 
Due to the spin-dependent selection rules for the optical transitions, the photoionization from the spin-up and spin-down branches induced by circularly polarized radiation occurs with different rates. 
 This is shown by the  vertical arrow representing the ``photoionization'' of helical states, i.e. the depopulation of the Dirac cone and population of the excited bulk states.
The resulting imbalance of the edge states population leads to a net electric current $J_c$.

Note that the photoexcited carriers can also contribute to the photocurrent but this contribution is small due to a fast relaxation of the bulk carriers.  A series of (Bi$_{1-x}$Sb$_x$)$_2$Te$_3$ thin films were
tailored so that the Fermi energy $\varepsilon_{\rm F}$ ranged from the bulk conduction-band bottom to the valence-band top through the surface-Dirac cone, see three insets in Fig.~\ref{fig_ti1}(b). The circular photogalvanic current, indicating a flow of spin-polarized surface-Dirac electrons, shows a pronounced peak when $\varepsilon_{\rm F}$ is set near the Dirac point, in correlation with the behaviour of carrier mobility.}

\begin{figure}[t]
\centering
\includegraphics*[width=6.7cm]{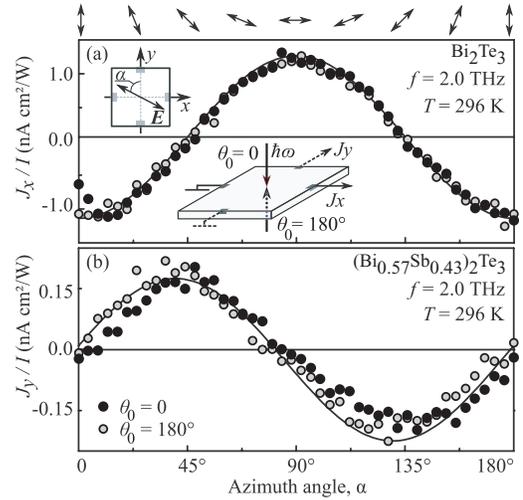}
\caption[]
{ 
{
%\color{blue}
({\bf a}) Photocurrent $J_x/I$ measured in a Bi$_2$Te$_3$ sample. ({\bf b}) Photocurrent $Jy/I$ measured in a (Bi$_{0.57}$Sb$_{0.43}$)$_2$Te$_3$ sample. Plots show the dependence of the photocurrent excited by normal incident radiation with $f$ = 2.0 THz on the azimuth angle $\alpha$. Angles of incidence $\theta_0$ = 0 and 180$^{\circ}$ correspond to the front and back excitations, respectively. Solid lines are best fits from the theory. Insets sketch the setup and the orientation of the electric field. Note that the photocurrent is probed in the directions coinciding with the principal axes of the trigonal system. After \cite{TopIns2add}} }
\label{fig_ti2}
\end{figure}

{
%\color{blue}
The orientational and polarization analysis of the photocurrents  in A$_2$B$_3$ TIs (A = Bi, Sb; B = Te, Se) and their alloys  is based on their bulk point-group symmetry D$_{3d}$ and the symmetry C$_{3v}$  of the surface states, the latter, of course, for the trigonal axis $C_3$ being normal to the sample surface. At normal incidence, the C$_{3v}$ point group forbids the circular PGE but allows the linear PGE.} 
{
%\color{blue}
This is why the helicity-dependent photocurrent is observed only at oblique angle of incidence ($\theta_0 \neq 0, 180^{\circ}$) \cite{TopIns4,TopIns1}, whereas the trigonal linear PGE is detected at any $\theta_0$ \cite{TopIns2,TopIns2add}.} 
{
%\color{blue}

The D$_{3d}$ group contains a center of space inversion and, thereby, forbids both the circular and linear photocurrents. However, a photon drag current can be generated in the centrosymmetric bulk as well. Under normal incidence of linearly polarized light, the azimuth-angle dependence of the photogenerated transverse current $J_x$ or $J_y$ is the same for the linear current of Dirac fermions and the drag current of bulk carriers. A straightforward way to distinguish the PGE response emerging from the surface states and the photon drag effect provides experiments with reversed direction of the light propagation: the former remains unchanged while the drag current has opposite directions in these two geometries.  Inset in Fig.~\ref{fig_ti2}(a) indicates two normal geometries of incidence, $\theta_0=0$ and $\theta_0 = 180^{\circ}$. One can see that the photocurrents measured under the front and back illumination coincide which clearly demonstrates that, at normal incidence, the photon drag yields a minor contribution to the total photocurrent. At oblique incidence, however, in particular at large angles of incidence, the linear photocurrent can be outweighed by the photon drag effect} 
{
%\color{blue} 
in the surface states or in the bulk \cite{TopIns2add}.} 

\begin{figure}[t]
\centering
\includegraphics*[width=8.5cm]{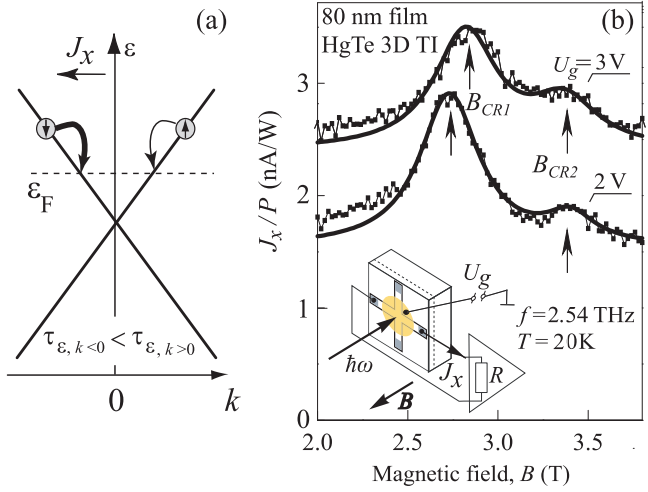}
% If not, use
%\picplace{5cm}{2cm} % Give the correct figure height and width in cm
\caption[]
{ 
{
%\color{blue}
({\bf a}) Microscopic origin of the cyclotron resonance induced spin photocurrent
in 3D TI materials. ({\bf b}) Magnetic field dependence of the photocurrent excited by circularly polarized THz radiation in a strained 80-nm HgTe 3D film. The numbers indicate applied gate voltages
(Fermi level remains in the band gap). Inset shows the experimental geometry. The data for $V_g = 2\ {\rm V}$ are multiplied by 4.} 
{
%\color{blue} 
After \protect\cite{TopIns3}.} }
\label{fig_ti3}
\end{figure}
{
%\color{blue}
In 3D TIs based on strained HgTe films the strain opens a gap in the otherwise gapless HgTe which, together with the high quality of the material, allows one to obtain insulation in the bulk and study surface charge transport only \cite{Bruene2011,Kozlov2014}. Figure \ref{fig_ti3} represents the observation of cyclotron resonance
induced photocurrents generated in the surface states of a strained HgTe film  sandwiched between thin Cd$_{0.65}$Hg$_{0.35}$Te layers acting as capping and buffer layers \cite{TopIns3}. Curved arrows in Fig.~\ref{fig_ti3}(a) indicate processes of carrier scattering. Their different thicknesses depict different scattering rates for the states with oppositely oriented spins due to an asymmetric correction to these rates caused by the mixing of electron states by the magnetic field. 

The asymmetry of carrier scattering in the momentum space leads to an electric current generation. Since  the electron spin orientation in a topological surface state is locked to its momentum the observed photocurrents are spin polarized and accompanied by the emergence of a macroscopic surface spin polarization. The model of the photocurrent formation is supported by complimentary measurements of the radiation transmission and magneto-transport, see for details Ref.~\cite{TopIns3}.

The excitation of the sample with right-handed circularly polarized radiation ($\sigma^+$) and sweep of the magnetic field leads to two resonant dips at positive magnetic fields, Fig.~\ref{fig_ti3}(b). Upon changing the radiation helicity from $\sigma^+$ to $\sigma^-$ the dips $B_{\rm CR1}$ and $B_{\rm CR2}$ appear at negative magnetic fields. For linearly polarized radiation, the resonances are observed for both magnetic field directions. These findings are clear signs that the absorption of radiation is caused by cyclotron resonance of electrons.
From the dip position $B_{\rm CR}$ one can determine the corresponding cyclotron mass as follows
\begin{equation}
m_{\rm CR} = \frac{|e| B_{\rm CR}}{2 \pi f c}\: , 
\label{eq_1}
\end{equation}
where $f$ is the light frequency. At $T$ = 40 K the masses $m_{\rm CR}$ = 0.028$m_0$ and 0.035$m_0$ are obtained for dips at $B_{\rm CR}$ = 2.6 and 3.35 T, respectively. The two values of $m_{\rm CR}$ correspond to the top and bottom interfaces, they are different because of a built-in electric field. It is very important to stress that the cyclotron resonance of the bulk carriers is expected at substantially higher magnetic fields corresponding to a mass of about 0.07$m_0$. }

\section{Circular photogalvanic effect in Weyl semimetals}
In the theoretical works \cite{Moore,Patrick,Koenig,spivak} the CPGE is studied in the Weyl semimetals. It is established that the contribution of each Weyl node to the circular photocurrent takes the universal form \cite{Moore}
\begin{equation} \label{Gamma00}
{\bm j} = {\cal C} \Gamma_0\tau_p {\rm i}\left( {\bm E} \times {\bm E}^* \right)\:,
\end{equation}
where $\Gamma_0 = \pi e^3 /3 h^2$, $e$ is the electron charge, $h$ is the Planck constant, ${\cal C} = \pm 1$ is the chirality (or topological charge) of the node, $\tau_p$ is the electron momentum relaxation time. The universality of Eq.~(\ref{Gamma00}) means that, with the exception of the factor $\tau_p$, the right-hand side of this equation contains a numerical factor $\pi/3$ and the world constants $e$ and $h$. Ching-Kit Chan et al. \cite{Patrick} have considered a pair of Weyl nodes with opposite chiralities. They have shown that the contributions to the circular photocurrent do not cancel each other provided that, in addition to the terms $A_{\alpha \beta} \sigma_{\alpha} k_{\beta}$,  the effective electron Hamiltonian contains the tilt term  ${\bm a}\cdot {\bm k}$ with the vector ${\bm a}$ different in the different nodes where ${\bm k}$ is the electron wave vector referred to the node ${\bm k}_W$. However, in this case the equation for the photocurrent loses its universality. Recently Qiong Ma et al. \cite{ExpLee} have observed a circular photocurrent in the TaAs crystal under excitation by CO$_2$ laser radiation.

Golub et al. \cite{spivak} have analyzed how the presence of a reflection plane in the point-symmetry group of a gyrotropic crystal affects the CPGE and discussed the influence of a magnetic field on the photocurrents in Weyl gyrotropic semimetals. A particular attention is paid to the crystal classes 
C$_{4v}$ and C$_{2v}$. The effective electron Hamiltonian near the Weyl nodet ${\bm k}_W$ is given  in the form
\begin{equation} \label{Hamilt}
{\cal H} = {\bm d}({\bm k})\cdot {\bm \sigma} + d_0({\bm k}) \sigma_0\:,
 \end{equation}
where $\sigma_0$ is the identity matrix of dimension 2, $d_l({\bm k})$ $(l=0,x,y,z)$ are functions whose expansions in powers of ${\bm k}$ contain no terms of the zero order. The eigenenergies of this Hamiltonian take on the values $E_{\pm, {\bm k}} = d_0({\bm k}) \pm d({\bm k})$, where $d(\bm k) = |{\bm d}(\bm k)|. $ Hereafter the energy is referred to the electron energy at the Weyl point. Usually, for simplicity only linear terms are taken into account in $d_0({\bm k})$: $d_0({\bm k}) = {\bm a} \cdot {\bm k}$, where ${\bm a}$ is some vector. This term describes the tilt of the Weyl cone. 

Under direct optical transitions in the vicinity of the ${\bm k}_W$ point, the following photocurrent is generated 
\begin{equation} \label{general}
{\bm j} = e \sum\limits_{\bm k} \tau_p \frac{2}{\hbar} \frac{\partial d({\bm k})}{\partial {\bm k}} W_{+-}({\bm k})\:,
\end{equation}
where the rate of optical transitions per unit volume per unit time is given by
 \begin{equation} \label{W+-}
W_{+-} = \frac{2 \pi}{\hbar} \left\vert M_{+-} \right\vert^2 F({\bm k})  \delta \left( 2d - \hbar \omega \right)\:,
\end{equation}
$M_{+-}$ is the matrix element of the optical transition that does not depend on $d_0({\bm k})$, 
\begin{equation} \label{tilt}
F({\bm k}) =  f \left(E_{-, {\bm k}}\right) - f \left(E_{+, {\bm k}} \right)\:,
\end{equation}
$f \left( E \right)$ is the equilibrium Fermi-Dirac distribution function. The factor 2 in Eq.~(\ref{general}) takes into account the contributions to the current from the photoelectrons and photoholes. Similarly to Ref.~\cite{Moore}, one can show that, under the circularly polarized excitation, the 
polarization-dependent contribution to the square modulus of the matrix element is proportional to the Berry curvature
\begin{equation}
\left\vert  M_{+-} \right\vert^2_{\rm circ} = \frac{2 e^2 d^2}{\left( \hbar \omega \right)^2} |{\bm E}|^2 {\bm \kappa}\cdot{\bm \Omega} \:,
\end{equation}
where ${\bm \kappa} = {\rm i} ({\bm e} \times {\bm e}^*)$, ${\bm e}$ is a unit polarization vector, the Berry curvature is related to the vector ${\bm d}({\bm k})$ by
\begin{equation}
\Omega_{j} = \frac{{\bm d}}{2d^3} \cdot \left(  \frac{\partial {\bm d}}{\partial k_{j+1}} \times \frac{\partial {\bm d}}{\partial k_{j+2}} \right) \:,
 \end{equation}
and the cyclic permutation of the indices is assumed. We note that, taking into account the energy conservation law, the argument $E_{\pm, {\bm k}}$ of the distribution function can be replaced by
$d_0({\bm k}) \pm \hbar \omega/2$. Therefore, for a fixed frequency $\omega$, the difference of occupation numbers (\ref{tilt}) is a function of the scalar $d_0({\bm k})$.

The structure of the second-rank pseudotensor ${\bm \gamma}$ in Eq.~(\ref{pge}) for all 18 gyrotropic classes is well known, see e.g. \cite{Sirotin}. The question arises what is the simplest form of the Hamiltonian (\ref{Hamilt}) which satisfies the two requirements: (i) it leads to a nonzero contribution of the node ${\bm k}_W$ to the $\gamma_{\alpha \beta}$ tensor which is allowed by the crystal symmetry group $F$, and (ii) this contribution does not disappear after the summation over the star of the vector ${\bm k}_W$.

First we consider {\it gyrotropic classes that do not contain reflection planes}, and take into account in the Hamiltonian (\ref{Hamilt}) only terms linear in ${\bm k}$: $d_{\alpha} = A_{\alpha \beta} k_{\beta}$, with the chirality of the node ${\bm k}_W$ equal to ${\rm sgn}\{ {\rm Det} ( \hat {\bm A} )\}$. In this case all the Weyl nodes obtained by the symmetry transformations are characterized by the same chirality. In the absence of a tilt, $d_0(\bm k) = 0$, the tensor ${\bm \gamma}$ is isotropic: the off-diagonal components are absent, while the diagonal components coincide and are equal to the contribution of a single node (\ref{Gamma00}) multiplied by the number of vectors $n$ in the star of the vector ${\bm k}_W$ if this star contains the vector $-{\bm k}_W$, and $2n$ if ${\bm k}_W$ and $-{\bm k}_W$ belong to different stars. The doubling is due to the symmetry to the time inversion which transforms the point ${\bm k}_W$ to $-{\bm k}_W$ preserving the chirality. The difference in the diagonal components allowed by the symmetry is obtained taking account of the tilt.

To calculate the off-diagonal components $\gamma_{\alpha \beta}$ in crystals of the symmetries 
C$_1$ and C$_{2}$, one needs to take into account the tilt with nonzero coefficients $a_{\alpha}$ and $a_{\beta}$. In the Hamiltonian with linear-${\bm k}$ terms, the off-diagonal components $\gamma_{xy} = -\gamma_{yx}$ in the classes C$_3$, C$_4$ and C$_6$ are not obtained, even with allowance for the tilt. 

{\it The gyrotropic classes containing reflection planes}, only linear-${\bm k}$ terms in the Hamiltonian are taken into account. The off-diagonal components of $\gamma_{\alpha \beta}$ in the groups C$_s$, C$_{2v}$, S$_{4}$ and the diagonal components $\gamma_{xx}= - \gamma_{yy}$ in the groups S$_4$,  D$_{2d}$ arise in the calculation with allowance for the tilt. In the groups 
C$_{3v}$, C$_{4v}$ and C$_{6v}$, nonzero off-diagonal components do not appear in the linear Hamiltonian model with an arbitrary tilt function $d_0({\bm k})$.

Thus, the six gyrotropic classes C$_n$, C$_{nv}$ ($n = 3,4,6$) stand apart from the rest: for them the components $\gamma_{xy}= - \gamma_{yx}$ can be obtained by adding, to the spin-dependent part of ${\cal H}$, terms of the higher (second or third) order in ${\bm k}$. In the ${\bm k}\cdot {\bm p}$ method, the nonlinear terms in ${\bm d}({\bm k})$ arise from the contribution of remote bands in the  perturbation theory and, therefore, can be considered as being small compared to the linear terms. 

One more effect specific for gyrotropic media is a magneto-induced photocurrent independent of the light polarization. In this effect,  in the linear in magnetic field approximation, the polar vector $-$ the photocurrent density $-$ is related with the axial vector $-$ the magnetic field. For example, the following currents are generated in crystals of the C$_{2v}$ symmetry \cite{spivak} 
\begin{eqnarray} \label{phenomen}
j_x &=& \left( S_{xx} B_x + S_{xy} B_y \right) |{\bm E}|^2 ,\\
j_y &=& - \left( S_{xx} B_y + S_{xy} B_x \right) |{\bm E}|^2\:, \nonumber
\end{eqnarray}
where the Cartesian coordinate system is chosen with the axis $z \parallel C_2$ and the axis $x$ composing the angle 45$^{\circ}$ with the reflection planes $\sigma_v$.

\section{Concluding Remarks}
The two main fields of study in the modern physics of semiconductors are transport phenomena and optical effects. Sometimes an impression arises that, founded on much the same
basis of tremendous successes achieved in technology, these fields are developing independently of each other. It is also true for the extensive studies of spin physics in semiconductors. One of the aims of this review is to show that the spin photogalvanics builds a solid bridge between the two fields and sets up a base for the reciprocation of ideas. Indeed, the spin-dependent photogalvanic effects, including charge and spin photocurrents, as well as the inverse effects allowing optical detection of
current-induced spin polarization need a thorough knowledge in both the transport physics and the polarized optical spectroscopy. As a result the different concepts supplement each other and provide a deeper insight in the spin-dependent microscopic processes.

Since the observation of the circular photogalvanic effect in GaAs quantum wells at the turn of the millennium~\cite{APL2000}, a coupling of photo-induced nonequilibrium spin to the directed electron motion has been demonstrated for a great variety of semiconductor low-dimensional systems. Experiments and theoretical developments demonstrated that spin-dependent optical excitation or spin-dependent relaxation of nonequilibrium carriers may cause an electric current as well as result in a pure spin current. The family of spin photogalvanics includes a number of phenomena such as the circular photogalvanic effect, spin-galvanic effect and its inversion, zero-bias spin separation, magneto-gyrotropic effect etc. The form of second-rank tensors $\bm{\gamma}, {\bm Q}, {\bm R}$ and the fourth-rank tensor ${\bm \Phi}$ describing these effects, i.e. the number of their nonzero components and possible relations between them, is fully determined by the space and time-inversion symmetry of the system. Therefore, the photocurrent measurements are a convenient tool for the determination or confirmation of the point-group symmetries of bulk crystals or nanoheterostructures. 

At present the spin photogalvanics has become a scientific measurement platform for the manifestations of various spin-dependent effects, such as the optical orientation of electronic spins, the Hanle effect, the exchange enhanced Zeeman effect in diluted magnetic semiconductors, the cyclotron resonance of topologically protected surface states, coherent Zitterbewegung  etc. As a result, the various band-structure and kinetic parameters can be readily and independently extracted from the photocurrent measurements.

As for the future work, the most challenging tasks are (i) the investigation of spin photogalvanics beyond perturbation limit at extremely high power optical excitation 
and (ii) the analysis of the photocurrent dynamics in time resolved experiments.  
Nonperturbative photogalvanics is expected to be characterized by a nontrivial behavior of the photocurrent on the light intensity and polarization state. This study should lead to the observation of the photovoltaic Hall effect at zero magnetic field and edge photocurrents caused by the nontrivial topology of Floquet states as 
well as would allow one to explore the energy spectrum reconstruction by intense 
THz electric fields due to formation of Floquet topological insulators and dressed states.

Time resolved experiments on spin photogalvanics under photoexcitation with femtosecond pulses, i.e. with the pulse duration being comparable with the free-carrier momentum relaxation time, would be desirable and informative for deep analysis of the individual photocurrent mechanisms like shift vs. ballistic  or spin vs. orbital effects.  Such study would also reveal a
great deal about the momentum, energy and spin relaxation of nonequilibrium photoexcited carriers. A technical realization of these challenging experiments becomes possible most recently and is, similarly to the Auston-switch, based on study of terahertz radiation emitted by current pulses by means of THz time domain spectroscopy.

%\printindex

\begin{thebibliography}{70}

\bibitem{Meier} F. Meier, B.P.~Zakharchenya (eds.), \textit{Optical Orientation} (North-Holland, Amsterdam, 1984)

\bibitem{Ivchenkobook2} E.L.~Ivchenko, \textit{ Optical
Spectroscopy of Semiconductor Nanostructures} (Alpha Science,
Harrow, 2005)

\bibitem{GanTrushSchl}   S.D. Ganichev, M.~Trushin, J. Schliemann, Spin Polarisation by Current, 
{
%\color{blue} 
in \textit{Handbook of Spin Transport and Magnetism},
eds. E.Y. Tsymbal, I. Zutic  (Chapman and Hall, 2011)}; arXiv:1606.02043 [cond-mat.mes-hall]





\bibitem{Ivchenko78p640} E.L.~Ivchenko and G.E.~Pikus,
Pis'ma Zh. Eksp. Teor. Fiz. {\bf 27}, 640 (1978); JETP Lett. {\bf 27}, 604 (1978)

\bibitem{Belinicher} V.I. Belinicher, Phys. Lett. A {\bf 66}, 213 (1978)

\bibitem{Asnin} V.M. Asnin, A.A. Bakun, A.M. Danishevskii, E.L.~Ivchenko, G.E.~Pikus, A.A.~Rogachev,
%\textit{Observation of a photo-emf that depends on the sign of the circular
%polarisation of the light,}
Pis'ma Zh. Eksp. Teor. Fiz. \textbf{28}, 80 (1978); JETP Lett. {\bf 28}, 74 (1978)

%\bibitem{Ch7Asnin78p74} V.M. Asnin, A.A. Bakun, A.M. Danishevskii,
%E.L. Ivchenko, G.E.~Pikus, and  A.A.~Rogachev,  Pis'ma Zh. \`{E}ksp. Teor. Fiz.
%\textbf{
%28}, 80-84 (1978) [JETP Lett. \textbf{ 28}, 74-77 (1978)].%-84  -77

\bibitem{sturman} B.I.~Sturman, V.M.~Fridkin, {\it The Photovoltaic and
Photorefractive Effects in Non-Centrosymmetric Materials} (Gordon
and Breach Science Publishers, Philadelphia, 1992)

\bibitem{APL2000}S.D. Ganichev, H. Ketterl, W. Prettl, E.L. Ivchenko, L.E. Vorobjev, Appl. Phys. Lett. {\bf 77}, 3146 (2000)

\bibitem{PRL01}S.D.~Ganichev, E.L. Ivchenko, S.N. Danilov, J.~Eroms,
W.~Wegscheider, D.~Weiss, W.~Prettl,
%\textit{Conversion of spin into directed electric current in quantum wells,}
Phys. Rev. Lett. \textbf{86}, 4358 (2001)%%-4361

\bibitem{Ganichev03p935} S.D.~Ganichev, W.~Prettl,
J. Phys.: Condens. Matter {\bf 15}, 935 (2003)

\bibitem{GanichevPrettl} S.D. Ganichev, W. Prettl, {\it Intense Terahertz Excitation of Semiconductors} (Oxford University Press, Oxford, 2006)


\bibitem{PRB03inv} S.D.~Ganichev, V.V.~Bel'kov, Petra~Schneider, E.L.~Ivchenko,
S.A.~Tarasenko, D.~Schuh, W.~Wegscheider, D.~Weiss, W.~Prettl,
%\textit{ Resonant inversion of circular photogalvanic effect in
%$n$-doped quantum wells,}
Phys.~Rev. B \textbf{68}, 035319 (2003) %6 pages    (cond-mat/0303054).

\bibitem{JETP06} V.~A. Shalygin, H.~Diehl, Ch.~Hoffmann, S.N.~Danilov, T.~Herrle, S.A.~Tarasenko, D.~Schuh,
Ch.~Gerl, W.~Wegscheider, W.~Prettl, S.D.~Ganichev,
%{\it Spin photocurrents and circular photon drag effect in (110)-grown structures},\\
Pis'ma Zh. Eksp. Teor. Fiz. \textbf{84}, 666 (2006); JETP Lett. \textbf{84}, 570 (2006)%666-672

\bibitem{C1} B. Wittmann, S.N. Danilov, V.V. Bel'kov, S.A. Tarasenko, E.G. Novik, H. Buhmann, C. Br{\"u}ne, L.W. Molenkamp, Z.D. Kvon, N.N. Mikhailov, S.A. Dvoretsky, N.Q. Vinh, A.F.G. van der Meer, B. Murdin, S.D. Ganichev, Semicond. Sci. Technol. {\bf 25}, 095005 (2010)

\bibitem{Dyakonov86p110} M.I. Dyakonov, V.Yu.~Kachorovskii,
%\textit{Spin relaxation of two-dimensional electrons in noncentrosymmetric semiconductors,}
Fiz. Tekh. Poluprovodn. \textbf{ 20}, 178 (1986); Sov. Phys. Semicond. \textbf{ 20}, 110 (1986)

\bibitem{A1Dresselhaus55p580} G.~Dresselhaus, %\textit{  Spin-orbit coupling %effects in zinc blende structures},
Phys. Rev. \textbf{ 100}, 580 (1955).%-586



\bibitem{Vasko79} F.T. Vasko, Pis'ma Zh. Eksp. Teor. Fiz., 30, 574 (1979); JETP Lett., 30, 541 (1979)

\bibitem{Bychkov84p78} Y.A. Bychkov, E.I.~Rashba,
%\textit{Properties of a 2D electron gas with lifted spectral degeneracy,}
Pis'ma Zh. Eksp. Teor. Fiz. \textbf{39}, 66
(1984); JETP Lett. \textbf{ 39}, 78 (1984)%%-69   -81

\bibitem{Winkler03} R.~Winkler,  \textit{Spin-Orbit Coupling Effects in Two-Dimensional Electron and Hole Systems}, Springer Tracts in Modern Physics, vol. 191  (Springer, Berlin, 2003)

\bibitem{Zawadzki2003pR1} W.~Zawadzki, P.~Pfeffer,
%\textit{ Spin splitting of subbands energies due to inversion asymmetry in
%semiconductor heterostructures},
Semicond. Sci. Technol. {\bf 19}, R1 (2004) %-R17

\bibitem{Krebs96p1829} O. Krebs, P. Voisin,
Phys. Rev. Lett. {\bf 77}, 1829 (1996)

\bibitem{roessler} U. R{\"o}ssler, J. Keinz, Solid State Commun. {\bf 121}, 313 (2002)

\bibitem{Nest08} M.O. Nestoklon, E.L. Ivchenko, P. Voisin, Phys. Rev. B {\bf 77}, 155328 (2008)
%Electric field effect on electron spin splitting in SiGe/Si quantum wells

\bibitem{PhysStatSolGolub}S.D. Ganichev, L.E. Golub, Phys. Stat. Sol. B {\bf 251}, 1801 (2014)

\bibitem{Belkov283p2003} V.V.~Bel'kov, S.D.~Ganichev, Petra~Schneider, C.~Back,
M.~Oestreich, J.~Rudolph, D.~H{\"a}gele, L.E.~Golub, W.~Wegscheider, W.~Prettl,
%\textit{ Circular photogalvanic effect at inter-band excitation in semiconductor quantum wells},
Solid State Commun. \textbf{128}, 283 (2003)%%-286

\bibitem{Bieler05} M.~Bieler, N.~Laman, H.M.~van~Driel, A.L.~Smirl,
%'Ultrafast spin-polarized electrical currents injected in a strained zinc blende semiconductors by single color pulses',
Appl. Phys. Lett. \textbf{86}, 061102 (2005)

\bibitem{Yang06} C.L.~Yang, H.T.~He, Lu~Ding, L.J.~Cui, Y.P.~Zeng, J.N.~Wang, W.K.~Ge,
%Spin photocurrent  and converse spin polarization induced in a InGaAs/InAlAs two-dimensional electron gas,
Phys. Rev. Lett. \textbf{96}, 186605 (2006)

\bibitem{Cho07} K.S.~Cho, Y.F.~Chen, Y.Q.~Tang, B.~Shen,
%Photogalvanic effects for interband absorption in AlGaN/GaN superlattices
Appl. Phys. Lett. \textbf{90}, 041909 (2007)

%
%TI:    Photogalvanic effects for interband absorption in AlGaN/GaN superlattices
%       AU:     Cho-KS; Chen-YF; Tang-YQ; Shen-B
%       SO:     Applied-Physics-Letters. 22 Jan. 2007; 90(4): 41909-1-3

\bibitem{YuChen2011}J.L. Yu, Y.H. Chen, C.Y. Jiang, Y. Liu, H. Ma, J. Appl. Phys. {\bf 109}, 053519 (2011)

\bibitem{YuChen2012}J. L. Yu, Y. H. Chen, Y. Liu, C.Y. Jiang, H. Ma, L. P. Zhu, Appl. Phys. Lett. {\bf 100}, 152110 (2012)
\bibitem{cpge2015}Jinling Yu, Shuying Cheng, Yunfeng Lai, Qiao Zheng, Laipan
Zhu, Yonghai Chen, Jun Ren, Opt. Express {\bf 23}, 027250 (2015)
%Temperature dependence of spin photocurrent spectra induced by Rashba- and Dresselhaus-type %circular photogalvanic effect at inter-band excitation in InGaAs/AlGaAs quantum wells
\bibitem{cpgeInN}Z. Zhang, R. Zhang, B. Liu, Z.L. Xie, X.Q. Xiu, P. Han, H. Lu, Y.D. Zheng, Y.H. Chen,
C.G. Tang, Z.G. Wang, Solid State Commun. {\bf 145} 159 (2008) 
\bibitem{giantRS} N. Ogawa, M.S. Bahramy, Y. Kaneko, Y. Tokura, Phys. Rev. B {\bf 90}, 125122 (2014)

\bibitem{Golub2003p235320} L.E. Golub,
%\textit{ Spin-splitting-induced photogalvanic effect in quantum wells},
Phys. Rev. B  \textbf{ 67}, 235320 (2003)%%7 pages

\bibitem{SpivakArx} E. Deyo, L.E. Golub, E.L. Ivchenko, B. Spivak, arXiv:0904.1917
%Semiclassical theory of the photogalvanic effect in non-centrosymmetric systems
\bibitem{Orenstein} J.E. Moore, J. Orenstein, Phys. Rev. Lett. {\bf 105}, 026805 (2010)


\bibitem{PRL02} S.D.~Ganichev, S.N. Danilov, V.V.~Bel'kov, E.L.~Ivchenko,
M.~Bichler, W.~Wegscheider, D.~Weiss, W.~Prettl,
%\textit{Spin-sensitive bleaching and monopolar spin orientation in quantum wells,}
Phys. Rev. Lett. \textbf{ 88}, 057401 (2002)%%4 pages

\bibitem{Schneider04p420}
P.~Schneider, J.~Kainz, S.D.~Ganichev, V.V.~Bel'kov, S.N.~Danilov,
M.M.~Glazov, L.E.~Golub, U.~R{\"o}ssler, W.~Wegscheider, D.~Weiss,
D.~Schuh, W.~Prettl,
%\textit{ Spin relaxation times of 2D
%holes from spin sensitive bleaching of inter-subband absorption},
J. Appl. Phys.  \textbf{ 96}, 420 (2004)%%-424

\bibitem{DyakPerel} M.I. Dyakonov, V.I. Perel, Zh. Eksp. Teor. Fiz. {\bf 60}, 1954 (1971); Sov. Phys. JETP {\bf 33}, 1053 (1971)


\bibitem{Ivchenko89p175} E.L.~Ivchenko, Yu.B.~Lyanda-Geller, G.E.~Pikus,
%\textit{Photocurrent in structures with quantum wells with an optical orientation of free carriers,}
Pis'ma Zh. Eksp. Teor. Fiz. \textbf{ 50}, 156  (1989); JETP Lett. \textbf{ 50}, 175 (1989)%%-158  -177

\bibitem{Nature02} S.D.~Ganichev, E.L.~Ivchenko,
V.V.~Bel'kov, S.A.~Tarasenko, M.~Sollinger, D.~Weiss,
W.~Wegscheider, W. Prettl,
%\textit{Spin-galvanic effect},
Nature (Lond.) \textbf{ 417}, 153 (2002) %%-156

\bibitem{Averkiev02pR271}  N.S.~Averkiev, L.E.~Golub, M.~Willander,
%\textit{Spin relaxation anisotropy in two-dimensional semiconductor systems},
J. Phys.: Condens. Matter \textbf{ 14}, R271 (2002)%%-R284


\bibitem{PRB03sge} S.D.~Ganichev, Petra~Schneider, V.V.~Bel'kov, E.L.~Ivchenko,
S.A.~Tarasenko, W.~Wegscheider, D.~Weiss, D.~Schuh, D.G.~Clarke,
 M.~Merrick, B.N.~Murdin, P.~Murzyn, P.J.~Phillips,  C.R.~Pidgeon,
 E.V.~Beregulin, W.~Prettl,
%\textit{ Spin galvanic effect due to optical spin orientation,}
Phys.~Rev. B  \textbf{68}, 081302 (2003)%%%%4 pages
%(cond-mat/0303193).
\bibitem{Elliot} R. J. Elliott, Phys. Rev. 96, 266 (1954).

\bibitem{Yafet} Y. Yafet, in Solid State Physics, vol. 14, ed. by F. Seitz, D. Turnbull (Academic, New York, 1963), p. 1.


\bibitem{GolubSGE}  L.E. Golub, Pis'ma Zh. Eksp. Teor. Fiz. \textbf{85}, 479 (2007);
JETP Lett. \textbf{ 85}, 393 (2007)

\bibitem{PhysRevB.70.155308.pdf}A.A. Burkov, A.S. N\'u$\tilde{n}$ez, A. H. MacDonald, Phys. Rev. B {\bf 70}, 155308 (2004)

\bibitem{PASPS02monop} S.A.~Tarasenko, E.L.~Ivchenko, V.V.~Bel'kov,
S.D.~Ganichev, D.~Schowalter, Petra~Schneider, M.~Sollinger,
W.~Prettl, V.M.~Ustinov,  A.E.~Zhukov,  L.E.~Vorobjev,
%``Monopolar optical orientation of electronic spins in semiconductors'',
J. Supercond.: Incorporating Novel Magn. \textbf{16}, 419 (2003)

\bibitem{Bakun84p1293} A.A.~Bakun, B.P.~Zakharchenya, A.A.~Rogachev, M.N.~Tkachuk,  V.G.~Fleisher,
Pis'ma ZhETF {\bf 40}, 464 (1984); Sov. JETP Lett. {\bf 40}, 1293 (1984)


\bibitem{Averkiev83p393} N.S.~Averkiev, M.I.~Dyakonov,
Fiz. Tekh. Poluprov. {\bf 17}, 629 (1983); Sov. Phys. Semicond. {\bf  17},  393 (1983)

\bibitem{Dyakonov71p144} M.I.~Dyakonov, V.I.~Perel',
Pis'ma ZhETF {\bf 13}, 206 (1971); Sov. JETP Lett. {\bf 13}, 144 (1971)


\bibitem{Ivchenko04p379} E.L.~Ivchenko, S.A.~Tarasenko,
%\textit{Optical orientation of electron spins in bulk semiconductors and heterostructures},
Zh. Eksp. Teor. Fiz. \textbf{ 126}, 426 (2004); JETP \textbf{ 99}, 379 (2004)%-434    -385

\bibitem{PRL04} S.D.~Ganichev, V.V.~Bel'kov, L.E.~Golub, E.L.~Ivchenko,
Petra~Schneider, S.~Giglberger, J.~Eroms, J.~De Boeck, G.~Borghs,
W.~Wegscheider, D.~Weiss, W.~Prettl,
 %\textit{ Experimental separation of Rashba and Dresselhaus
%spin-splittings in semiconductor quantum wells},
Phys. Rev. Lett. \textbf{ 92}, 256601 (2004) %%4 pages
\bibitem{PRB07gig} S.~Giglberger, L.E.~Golub,  V.V.~Bel'kov, S.N.~Danilov, D.~Schuh, Ch.~Gerl,
F.~Rohlfing,  J.~Stahl, W.~Wegscheider, D.~Weiss, W.~Prettl, S.D.~Ganichev,
%\\\textit{Rashba and Dresselhaus Spin-Splittings in Semiconductor Quantum Wells Measured by Spin Photocurrents},\\
Phys. Rev. B \textbf{75} 035327 (2007)

%\bibitem{Pikus1} F.G. Pikus, G.E. Pikus, Phys. Rev. B {\bf 51}, 16928 (1995)
%%Conduction-band splitting and negative magnetoresistance in A$_3$B$_5$ heterostructures
%\bibitem{Pikus2} W. Knap, C. Skierbiszewski, A. Zduniak, E. Litwin-Staszewska, D. Bertho, F. Kobbi, J.L. Robert, G.E. Pikus, F.G. Pikus, S.V. Iordanskii, V. Mosser, K. Zekentes, Yu.B. Lyanda-Geller, Phys. Rev. B {\bf 53}, 3912 (1996)
%%Weak antilocalization and spin precession in quantum wells

\bibitem{helix} M. Kohda, V. Lechner, Y. Kunihashi, T. Dollinger, P. Olbrich, C. Sch{\"o}nhuber, I. Caspers, V.V. Bel'kov, L.E. Golub, D. Weiss, K. Richter, J. Nitta, S.D. Ganichev, Phys. Rev. B {\bf 86}, 081306 (2012)
\bibitem{zitter} I. Stepanov, M. Ersfeld, A.V.~Poshakinskiy, M.~Lepsa,
E.L.~Ivchenko, S.A.~Tarasenko, B.~Beschoten, submitted

\bibitem{Vorobjev79p441} L.E.~Vorob'ev,
E.L.~Ivchenko, G.E.~Pikus, I.I.~Farbstein, V.A.~Shalygin, 
A.V.~Sturbin, Pis'ma Zh. Eksp. Teor. Fiz. {\bf 29}, 485
(1979); JETP Lett. {\bf 29}, 441 (1979)

\bibitem{Tellurium2012}V.A. Shalygin,  A.N. Sofronov, L.E. Vorob'ev, I.I. Farbshtein,  Fiz. Tverd. Tela {\bf 54},  2237 (2012); Phys. Solid State {\bf 54}, 2362 (2012)


\bibitem{Vasko} F.T.~Vasko and N.A. Prima, Fiz. Tverd. Tela {\bf
21}, 1734 (1979) [Sov. Phys. Solid State {\bf 21}, 994 (1979)].

\bibitem{levitov} L.S. Levitov, Yu.V. Nazarov, and G.M. Eliashberg, Zh. Eksp. Teor. Fiz. \textbf{88}, 229
(1985); Sov. Phys. JETP \textbf{61}, 133 (1985)


\bibitem{Aronov89p431} A.G. Aronov, Yu.B. Lyanda-Geller,
%\textit{Nuclear electric resonance and orientation of carrier
%spins by an electric field,}
Pis'ma Zh. Eksp. Teor. Fiz. \textbf{ 50}, 398 (1989); JETP Lett. \textbf{ 50}, 431
(1989)%%-400  -434

\bibitem{Edelstein89p233} V.M. Edelstein, Solid State Commun.~\textbf{ 73}, 233 (1990) %%-235
%\textit{Spin polarization of conduction
%electrons induced by electric current in two-dimensional
%asymmetric electron systems},


\bibitem{Aronov91p537} A.G. Aronov, Yu.B. Lyanda-Geller, G.E. Pikus,
%\textit{Spin polarization ofelectrons by an electric current,}
Zh Eksp. Teor. Fiz. \textbf{100}, 973 (1991); Sov. Phys. JETP \textbf{73}, 537
(1991)%%-400  -434

\bibitem{Chaplik} A.V. Chaplik, M.V. Entin, L.I. Magarill
%Spin orientation of electrons by lateral electric field in 2D system without inversion symmetry
Physica E \textbf{13}, 744 (2002)
%744 - 747

\bibitem{VaskoRai} F.T. Vasko, O.E. Raichev, {\it Quantum Kinetic Theory and
Applications} (Springer, New York, 2005).

\bibitem{TarJETP} S.A. Tarasenko, Pis'ma Zh. Eksp. Teor. Fiz. {\bf 84},
233 (2006); JETP Lett. {\bf 84}, 199 (2006)

\bibitem{Schlieman07} M.~Trushin, J. Schliemann,
%  .Anisotropic current-induced spin accumulation in the two-dimensional electron gas with spin-orbit coupling
Phys. Rev. B \textbf{75}, 155323 (2007)

\bibitem{Raichev} O.E. Raichev, Phys. Rev. B \textbf{75}, 205340 (2007)
\bibitem{Tarasenko2008}S.A. Tarasenko, Physica E {\bf 40}, 1614 (2008)
\bibitem{GolIvch} L.E. Golub, E.L. Ivchenko, Phys. Rev. B {\bf 84}, 115303 (2011)


\bibitem{Raimondi2016}D. Guerci, J. Borge, R. Raimondi, Physica E {\bf 82} 151 (2016)
%Physica E 82 , 151 (2016)

\bibitem{Gorini2017} C. Gorini, A. Maleki, Ka Shen, I.V. Tokatly, G. Vignale, and R. Raimondi,
%Theory of current-induced spin polarizations in an electron gas
arXiv:1702.04887v1
%  [cond-mat.mes-hall]  16 Feb 
(2017)


\bibitem{Ganichev04p0403641}S.D.~Ganichev, S.N.~Danilov, Petra~Schneider, V.V.~Bel'kov,
L.E.~Golub, W.~Wegscheider, D.~Weiss, W.~Prettl,
{
%\color{blue}
arXiv:0403641 (2004),  see also J. Magn. and Magn. Materials \textbf{300}, 127 (2006)}

\bibitem{Silov2004} A.Yu. Silov, P.A. Blajnov,  J.H. Wolter,
R.~Hey, K.H.~Ploog, N.S.~Averkiev,
%\textit{ Current-induced spin polarization at a single heterojunction},
Appl. Phys. Lett. \textbf{ 85}, 5929 (2004)%%-5931

\bibitem{Kato04p176601} Y.K. Kato, R.C. Myers, A.C. Gossard, D.D.~Awschalom,
%\textit{ Current-induced spin polarization in strained semiconductors},
Phys. Rev. Lett. \textbf{ 93}, 176601 (2004)%%%%4 pages

\bibitem{Silov04} A.Yu. Silov, P.A. Blajnov, J.H. Wolter,  R. Hey,
K.H. Ploog, and N.S. Averkiev, in \textit{Proc. 13th Int. Symp.
Nanostructures: Phys. and Technol.}, (St. Petersburg, Russia,
2005)

\bibitem{Sih05isge} V.~Sih, R.C. Myers, Y.K. Kato, W.H.~Lau, A.C. Gossard, D.D.~Awschalom,
%Imaging work in (110) QW
Nat. Phys. \textbf{1}, 31 (2005)

\bibitem{Stern06isge} N.P.~Stern, S.~Ghosh, G.~Xiang, M.~Zhu, N.~Samarth, D.D.~Awschalom,
%``Current-induced polarization and the spin Hall effect at room temperature'',
Phys. Rev. Lett. \textbf{97}, 126603 (2006)%cond-mat/0607288 (2006).
\bibitem{Beschoten} S. Kuhlen, K. Schmalbuch, M. Hagedorn, P. Schlammes, M. Patt, M. Lepsa,
G. G{\"u}ntherodt, and B. Beschoten, Phys. Rev. Lett. {\bf 109}, 146603 (2012)

\bibitem{Awschalom} B.M. Norman, C.J. Trowbridge, D.D. Awschalom, V. Sih, Phys. Rev. Lett. {\bf  112}, 056601 (2014)
%Current-Induced Spin Polarization in Anisotropic Spin-Orbit Fields

\bibitem{streaming}L.E. Golub, E.L. Ivchenko, New J. Phys. {\bf 15}, 125003 (2013)

\bibitem{hopping}
%Electrical Spin Orientation, Spin-Galvanic, and Spin-Hall Effects in Disordered Two-Dimensional Systems, 
D.S. Smirnov and L.E. Golub, Phys. Rev. Lett. \textbf{118}, 116801  (2017)
%D. S. Smirnov, L. E. Golub, 33rd Internat. Conf. Phys. Semicond. (Beijing, China, 2016) Abstracts, p. 238

\bibitem{spintronicbook02} D.D. Awschalom, D. Loss, N. Samarth (eds.), {\it Semiconductor
Spintronics and Quantum Computation} (Springer, Berlin, 2002)

\bibitem{Bhat} R.D.R.~Bhat, F.~Nastos, A.~Najmaie,
J.E.~Sipe, Phys. Rev. Lett. {\bf 94}, 096603 (2005)

\bibitem{Tarasenko05p292} S.A. Tarasenko, E.L. Ivchenko,
%\textit{ Pure spin photocurrents in low-dimensional structures,}
Pis'ma Zh. Eksp. Teor. Fiz. \textbf{ 81}, 292 (2005); JETP Lett. \textbf{ 81}, 231 (2005)%%-296  -235

\bibitem{Zhao} H.~Zhao, X.~Pan, A.L.~Smirl, R.D.R. Bhat, A. Najmaie, J.E. Sipe, H.M.~van~Driel,
Phys. Rev. B {\bf 72} 201302 (2005)

\bibitem{purespin2}Bin Zhou, Shun-Qing Shen, Phys. Rev. B {\bf 75}, 045339 (2007)

\bibitem{Ganichev06zerobias} S.D.~Ganichev, V.V.~Bel'kov, S.A.~Tarasenko, S.N.~Danilov, S.~Giglberger,
Ch.~Hoffmann, E.L.~Ivchenko, D.~Weiss, W.~Wegscheider,  Ch.~Gerl,
D.~Schuh, J.~Stahl, J.De~Boeck, G.~Borghs, W.~Prettl,
Nat. Phys. \textbf{2}, 609 (2006)

\bibitem{PRBSiGe} S.D.~Ganichev, S.N. Danilov, V.V.~Bel'kov, S.~Giglberger, S.A. Tarasenko, E.L.~Ivchenko, D.~Weiss, W. Jantsch, 
F. Sch{\"a}ffler, D. Gruber, W.~Prettl,
 %\textit{ Pure spin currents induced by spin-dependent scattering
%processes in SiGe quantum well structures},
Phys. Rev. B \textbf{75}, 155317 (2007)

\bibitem{TarICPS} S.A. Tarasenko, E.L. Ivchenko, Proc. ICPS-28 (Vienna, 2006),
AIP Conf. Proc. {\bf 893}, 1331 (2007); cond-mat/0609090

\bibitem{voisin} E.L. Ivchenko, A.A. Toropov, P. Voisin, Fiz. Tverd. Tela {\bf 40}, 1925 (1989); Phys. Solid State \textbf{40}, 1748 (1998)

\bibitem{Khurgin} J.B.~Khurgin, Phys. Rev. B {\bf 73}, 033317 (2006)

\bibitem{Jancu} J.-M.~Jancu, R.~Scholz, E.A.~de~Andrada~e~Silva,
G.C.~La~Rocca, Phys. Rev. B {\bf 72}, 193201 (2005)


\bibitem{Stevens02p4382}M.J.~Stevens, A.L.~Smirl, R.D.R.~Bhat, J.E.~Sipe, H.M.~van~Driel,
J. Appl. Phys. {\bf 91}, 4382 (2002)

\bibitem{Stevens2003} M.J. Stevens, A.L. Smirl, R.D.R. Bhat, A. Najimaie, J.E. Sipe,  H.M.~van~Driel,
%'Quantum Interference Control of Ballistic Pure Spin Currents in Semiconductors',
Phys. Rev. Lett. \textbf{90}, 136603 (2003)

\bibitem{Entin89p664} M.V.~Entin, Fiz. Tekh. Poluprov. {\bf 23}, 1066 (1989);
Sov. Phys. Semicond. {\bf 23}, 664 (1989)

\bibitem{Belkov05p3405} V.V.~Bel'kov, S.D.~Ganichev, E.L.~Ivchenko,
S.A.~Tarasenko,  W.~Weber, S.~Giglberger,  M.~Olteanu, P.~Tranitz,
S.N.~Danilov, Petra~Schneider, W.~Wegscheider, D.~Weiss, 
W.~Prettl,
%\textit{ Magneto-gyrotropic photogalvanic effect in
%semiconductor quantum wells},
J. Phys.: Condens. Matter \textbf{ 17}, 3405 (2005) 

{
%\color{blue} 
\bibitem{Belkovbook} V.V. Bel'kov, S.D. Ganichev,
\textit{Zero-bias spin separation}, in \textit{Handbook of Spintronic Semiconductors}, 
eds. W.M.~Chen and I.A.~Buyanova (Pan Stanford Publishing 2010)

\bibitem{Olbrich2012}  P.\,Olbrich, C.\,Zoth, P.\,Vierling, K.-M.\,Dantscher,
G.V.\,Budkin, S.A.\,Tarasenko, V.V.\,Bel'kov, D.A.\,Kozlov,  Z.D.\,Kvon,
N.N.\,Mikhailov, S.A.\,Dvoretsky, S.D.\,Ganichev,
%\textit{Giant photcurrent in a Dirac fermion system at cyclotron resonance}\\
Phys. Rev. B  \textbf{87}, 235439 (2013) }

\bibitem{67} V.V. Bel'kov, P. Olbrich, S.A. Tarasenko, D. Schuh, W. Wegscheider, T. Korn, C. 
Sch{\"u}ller, D. Weiss, W. Prettl, S.D. Ganichev, Phys. Rev. Lett. {\bf 100}, 176806 (2008)

\bibitem{spivakx} E.L. Ivchenko, B. Spivak, Phys. Rev. B {\bf 66}, 155404
(2002)
\bibitem{Torbital} S.A.~Tarasenko, Pis'ma Zh. Eksp. Teor. Fiz. \textbf{85}, 216 (2007); JETP Lett. \textbf{85}, 182 (2007)
\bibitem{Orbital2010} J. Karch, S.A. Tarasenko, P. Olbrich, T. Sch{\"o}nberger,
C. Reitmaier, D. Plohmann, Z.D. Kvon, S.D. Ganichev, J. Phys.: Condens. Matter {\bf 22}, 355307 (2010)
\bibitem{Taras2011}S.A. Tarasenko, Phys. Rev. B  \textbf{83}, 035313  (2011)
\bibitem{TarasNature} C. Drexler,	S.A. Tarasenko,	P. Olbrich,	J. Karch, M. Hirmer,	F. M{\"u}ller,	M. Gmitra, J. Fabian,	R. Yakimova,	S. Lara-Avila,	S. Kubatkin,	M. Wang,	R. Vajtai,	P. M. Ajayan,	J. Kono, and S.D. Ganichev,  Nature Nanotechn. {\bf 8}, 104 (2013)
\bibitem{nanowireSi} S. Dhara, E.J. Mele, R. Agarwal, Science {\bf 349}, 726 (2015)
%Voltage tunable circular photogalvanic effect in Si Nanowires 
\bibitem{Falko}N. Kheirabadi, E. McCann, V.I. Fal'ko,  arXiv:1606.08234 
\bibitem{circSi}P. Olbrich, S.A. Tarasenko, C. Reitmaier, J. Karch, D. Plohmann, Z.D. Kvon, S.D. Ganichev, Phys. Rev. B {\bf 79}, 121302 (2009)
\bibitem{orbital0} E.L. Ivchenko, S.D. Ganichev, Pis'ma Zh. Eksp. Teor. Fiz. \textbf{93}, 752 (2007); JETP Lett. \textbf{93}, 673 (2011)
\bibitem{orbital1}J. Karch, P. Olbrich, M. Schmalzbauer, C. Zoth, C. Brinsteiner, M. Fehrenbacher, U. Wurstbauer, M.M. Glazov, S.A. Tarasenko, E.L. Ivchenko, D. Weiss, J. Eroms, R. Yakimova, S. Lara-Avila, S. Kubatkin, S.D. Ganichev, Phys. Rev. Lett. {\bf 105}, 227402 (2010)
\bibitem{orbital2} J. Karch, S.A. Tarasenko, E.L. Ivchenko, J. Kamann, P. Olbrich1 M. Utz, Z.D. Kvon, S.D. Ganichev, Phys. Rev. B {\bf 83}, 121312 (2011)
\bibitem{orbital3}  C. Drexler,	S.A. Tarasenko,	P. Olbrich,	J. Karch, M. Hirmer,	F. M{\"u}ller,	M. Gmitra, J. Fabian,	R. Yakimova,	S. Lara-Avila,	S. Kubatkin, M. Wang, R. Vajtai,	P.M. Ajayan, J. Kono, S. D. Ganichev, Nat.  Nanotechn.  {\bf 8}, 104 (2013)
%Magnetic ratchet effect in bilayer graphene
%\bibitem{Knippels99p1578}  G.M.H.~Knippels, X.~Yan, A.M.~MacLeod, W.A.~Gillespie,
%M.~Yasumoto, D.~Oepts and A.F.G.~van der~Meer,
%%\textit{Generation and complete electric-field characterization of intense ultrashort
%%tunable far-infrared laser pulses,}
%\textit{Phys. Rev. Lett.} \textbf{ 83},
%1578 (1999).%%-1581
\bibitem{gFactorSign}V. Lechner, L. E. Golub, F. Lomakina, V.V. Bel'kov, P. Olbrich, S. Stachel, I. Caspers, M. Griesbeck, M. Kugler, M. J. Hirmer, T. Korn, C. Sch{\"u}ller, D. Schuh, W. Wegscheider, S.D. Ganichev, Phys. Rev. B {\bf 83}, 155313 (2011)
\bibitem{diluted}P. Olbrich, C. Zoth, P. Lutz, C. Drexler, V.V. Bel'kov, Ya.V. Terent'ev, S.A. Tarasenko, A.N. Semenov, S.V. Ivanov, D.R. Yakovlev, T. Wojtowicz, U. Wurstbauer, D. Schuh, S.D. Ganichev, Phys. Rev. B {\bf 86}, 085310 (2012)

\bibitem{Hasan} M.Z. Hasan, C.L. Kane, Rev. Modern Phys. {\bf 82}, 3045 (2010)

\bibitem{Ando}Y. Ando, J. Phys. Soc. Japan {\bf  82}, 102001 (2013)



\bibitem{inverseTI}I. Garate, M. Franz, Phys. Lett. {\bf 104}, 146802 (2010)
%Inverse Spin-Galvanic Effect in the Interface between a Topological Insulator and a Ferromagnet

\bibitem{HosurBerry}P. Hosur, Phys. Rev. B {\bf 83}, 035309 (2011)
%Circular photogalvanic effect on topological insulator surfaces:
%Berry-curvature-dependent response

\bibitem{TopIns4}J.W. McIver, D. Hsieh, H. Steinberg, P. Jarillo-Herrero, N. Gedik, Nat. Nanotechnol. {\bf 7}, 96 (2012)
%Control over topological insulator photocurrents with light polarization

\bibitem{Ultrathin} Quan Sheng Wu, Sheng Nan Zhang, Zhong Fang, Xi Dai, Physica E {\bf 44}, 895 (2012)
%Photogalvanic in ultrathin film of topological insulator

\bibitem{TopIns2}P. Olbrich, L.E. Golub, T. Herrmann, S.N. Danilov, H. Plank, V.V. Bel'kov, G. Mussler, Ch. Weyrich, C.M. Schneider, J. Kampmeier, D. Gr{\"u}tzmacher, L. Plucinski, M. Eschbach, S.D. Ganichev, Phys. Rev. Lett. {\bf 113}, 096601 (2014)
%Room-Temperature High-Frequency Transport of Dirac Fermions in Epitaxially Grown
%Sb2Te3- and Bi2Te3-Based Topological Insulators

\bibitem{TopIns5} Junxi Duan, Ning Tang, Xin He, Yuan Yan, Shan Zhang, Xudong Qin, Xinqiang Wang, Xuelin Yang, Fujun Xu, Yonghai Chen, Weikun Ge, Bo Shen, Sci. Rep. {\bf 4}, 4889
(2014)
%Identification of Helicity-Dependent Photocurrents from Topological Surface
%States in Bi2Se3 Gated by Ionic Liquid

\bibitem{TItheory}V. Kaladzhyan, P.P. Aseev, S.N. Artemenko, Phys. Rev. B {\bf 92}, 155424 (2015)
%Photogalvanic effect in the HgTe/CdTe topological insulator due to edge-bulk optical transitions

\bibitem{TopIns3dan}K.-M. Dantscher, D.A. Kozlov, P. Olbrich, C. Zoth, P. Faltermeier, M. Lindner, G.V. Budkin, S. A. Tarasenko, V.V. Bel'kov, Z. D. Kvon, N.N. Mikhailov, S.A. Dvoretsky, D. Weiss, B. Jenichen, S.D. Ganichev, Phys. Rev. B {\bf 92}, 165314 (2015)
%Cyclotron-resonance-assisted photocurrents in surface states of a three-dimensional topological
%insulator based on a strained high-mobility HgTe film

\bibitem{Ultrafast} Christoph Kastl, Christoph Karnetzky, Helmut Karl, A.W. Holleitner, Nat. Commun. {\bf 6}, 6617 (2015)
%Ultrafast helicity control of surface currents in topological insulators with near-unity fidelity

\bibitem{TopIns1}K.N. Okada, N. Ogawa, R. Yoshimi, A. Tsukazaki, K.S. Takahashi,
M. Kawasaki, Y. Tokura, Phys. Rev. B {\bf 93}, 081403 (2016)
%Enhanced photogalvanic current in topological insulators via Fermi energy tuning

\bibitem{TopIns2add} H. Plank, L.E. Golub, S. Bauer, V.V. Bel'kov, T. Herrmann, P. Olbrich, M. Eschbach, 
L. Plucinski, C.M. Schneider, J. Kampmeier, M. Lanius, G. Mussler, D. Gr{\"u}tzmacher, S.D. Ganichev, 
Phys. Rev. B {\bf 93}, 125434 (2016)
%Photon drag effect in (Bi1?xSbx)2Te3 three-dimensional topological insulators

\bibitem{TopIns3} H. Plank, S.N. Danilov, V.V. Bel'kov, V.A. Shalygin, J. Kampmeier, M. Lanius, G. Mussler, D. Gr{\"u}tzmacher, and S.D. Ganichev,
%%\textit{Opto-Electronic Characterization of Three Dimensional Topological Insulators}\\
J. Appl. Phys. \textbf{120}, 165301 (2016)

%\color{blue}
\bibitem{TopIns2add2}  
 K.-M. Dantscher, D.A. Kozlov, M.-T. Scherr, S. Gebert, J. B{\"a}renf{\"a}nger, V.V. Bel'kov, 
S.A. Tarasenko, M.V. Durnev, N.N. Mikhailov, S.A. Dvoretskii, Z.D. Kvon, S.D. Ganichev,
\textit{Chiral edge photogalvanic current induced in 2D HgTe topological insulators},
Int. Conf. on New Trends in Topological Insulators, W{\"u}rzburg (2016)


\bibitem{Bruene2011} C. Br{\"u}ne, C.X. Liu, E.G. Novik, E.M. Hankiewicz, H. Buhmann, Y.L. Chen, X.L. Qi, Z.X. Shen, S.C. Zhang, L.W. Molenkamp, Phys. Rev. Lett. {\bf 106}, 126803 (2011)

\bibitem{Kozlov2014} D.Z. Kozlov, Z.D. Kvon, E.B. Olshanetsky, N.N. Mikhailov, S.D. Dvoretsky, D. Weiss, Phys. Rev. Lett. {\bf 112}, 196801 (2014)
%%%%%%%%%%%%%%%%%%%%%%%










\bibitem{Moore} F.~de~Juan, A.G. Grushin, T. Morimoto, J.E.~Moore, Nat. Commun. {\bf 8}, 15995 (2017).
\bibitem{Patrick} Ching-Kit Chan, N.H. Lindner, G. Refael, P.A. Lee, Photocurrents in Weyl semimetals, Phys. Rev. B \textbf{95}, 041104 (2017).
\bibitem{Koenig} E.J. K{\"o}nig, H.-Y. Xie, D.A. Pesin, A. Levchenko,  	Phys. Rev. B {\bf 96}, 075123 (2017).
\bibitem{spivak} L.E. Golub, E.L. Ivchenko, B.Z. Spivak, Pis'ma Zh. Eksp. Teor. Fiz. \textbf{105}, 744 (2007); JETP Lett. {\bf 105}, 782 (2017).
\bibitem{ExpLee}  Q. Ma, S.-Y. Xu, C.-K. Chan, C.-L. Zhang, G. Chang, Y. Lin, W. Xie, T. Palacios, H. Lin, S. Jia, P.A. Lee, P. Jarillo-Herrero, N. Gedik, 
Nature Phys. \textbf{13}, 842 (2017).
%arXiv:1705.00590v1.
\bibitem{Sirotin} Yu.I. Sirotin, M.P. Shaskolskaya, Fundamentals of Crystal Physics (Nauka, Moscow, 1975; Mir, Moscow, 1982).

\end{thebibliography}
\end{document}